\documentclass[12pt]{article}
\usepackage{bbm}
\usepackage{epsfig}
\usepackage{array}
\usepackage{float}
\usepackage{dsfont}

\usepackage{a4}

\usepackage{a4wide}

\def\be{\begin{equation}}
\def\ee{\end{equation}}
\def\gs{\mathrel{
   \rlap{\raise 0.511ex \hbox{$>$}}{\lower 0.511ex \hbox{$\sim$}}}}
\def\ls{\mathrel{
   \rlap{\raise 0.511ex \hbox{$<$}}{\lower 0.511ex \hbox{$\sim$}}}}

\newcommand{\ba}{\begin{array}{c}}
\newcommand{\baz}{\begin{array}{cc}}
\newcommand{\bad}{\begin{array}{ccc}}
\newcommand{\bea}{\begin{equation} \begin{array}{c}}
\newcommand{\eea}{ \end{array} \end{equation}}
\newcommand{\ea}{\end{array}}
\newcommand{\D}{\displaystyle}

\newcommand{\sss}{\sin^2 \theta_{12}}

\begin{document}

\title{\vspace{-3cm}
\hfill {\small MPP--2007--27}\\
\vspace{-0.3cm}
\hfill {\small SISSA 44/2007/EP}\\
\vspace{-0.3cm} 
\hfill {\small arXiv:0706.2975} 
\vskip 0.4cm
\bf 
$U_{\rm PMNS} = U_\ell^\dagger \, U_\nu$ 
}
\author{
K.~A.~Hochmuth$^a$\thanks{email: \tt hochmuth@mppmu.mpg.de}~~,~~
S.~T.~Petcov$^b$\thanks{Also at: Institute of
Nuclear Research and Nuclear Energy,
Bulgarian Academy of Sciences, 1784 Sofia, Bulgaria.
}~\mbox{}
~~and~
W.~Rodejohann$^c$\thanks{email: \tt werner.rodejohann@mpi-hd.mpg.de} 
\\\\
{\normalsize \it $^a$Max--Planck--Institut f\"ur Physik
 (Werner-Heisenberg-Institut),}\\
{\normalsize \it F\"ohringer Ring 6, 
D--80805 M\"unchen, Germany}\\ \\ 
{\normalsize \it $^{b}$Scuola Internazionale Superiore di Studi Avanzati,}\\
{\normalsize \it Via Beirut 2--4, I--34014 Trieste, Italy}\\
{\normalsize and}\\
{\normalsize \it Istituto Nazionale di Fisica Nucleare,}\\
{\normalsize \it Sezione di Trieste, I--34014 Trieste, Italy}
\\ \\
{\normalsize \it $^c$Max--Planck--Institut f\"ur Kernphysik,}\\
{\normalsize \it Postfach 10 39 80, D--69029 Heidelberg, Germany}
}
\date{}
\maketitle
\thispagestyle{empty}
\vspace{-0.8cm}
\begin{abstract}
\noindent 
We consider corrections to vanishing $U_{e3}$ and maximal atmospheric 
neutrino mixing originating from the relation 
$U = U_\ell^\dagger \, U_\nu$, where $U$ is the PMNS mixing matrix and 
$U_\ell$ $(U_\nu)$ is  associated with the diagonalization 
of the charged lepton (neutrino) mass matrix. 
We assume that in the limit of $U_\ell$ or $U_\nu$ 
being the unit matrix, 
one has $U_{e3} = 0$ and $\theta_{23} = \pi/4$,
while the  solar neutrino mixing angle 
is a free parameter.
Well-known special cases of 
the indicated scenario 
are the bimaximal and tri-bimaximal mixing 
schemes. If  $U_{e3} \neq 0$ and $\theta_{23} \neq \pi/4$
due to corrections from the charged leptons, 
$|U_{e3}|$ can be sizable (close to the existing upper limit) 
and we find that the value of the solar neutrino mixing angle 
is linked to the magnitude of CP violation in neutrino oscillations.
In the alternative case of the neutrino sector correcting 
$U_{e3} = 0$ and $\theta_{23} = \pi/4$, we obtain a generically smaller
$|U_{e3}|$ than in the first case. 
Now the magnitude of CP violation in neutrino 
oscillations is connected 
to the value of the atmospheric 
neutrino mixing angle $\theta_{23}$. 
We find that both cases are in 
agreement with present observations. We also 
introduce parametrization independent ``sum-rules'' 
for the oscillation parameters.
  
\end{abstract}

\newpage

\section{\label{sec:intro}Introduction}

The low energy neutrino mixing implied by the neutrino oscillation
data can be described by the Lagrangian (see, e.g., \cite{reviews}) 
\be 
\label{eq:L}
{\cal L} = 
- \frac{g}{\sqrt{2}} \,
\overline{\ell_L} \, \gamma^\mu \, \nu_L \, W_\mu 
- \frac 12 \, \overline{\nu_{R}^c} \, m_{\nu} \,  \nu_{L} 
-  \overline{\ell_R} \, m_\ell \, \ell_L  + h.c., 
\ee
%
which includes charged lepton and Majorana neutrino 
mass terms. 
When diagonalizing the neutrino and charged lepton 
mass matrices via 
$m_\nu = U_\nu^\ast \, m_\nu^{\rm diag} \,U_\nu^\dagger$ 
and $m_\ell = V_\ell \, m_\ell^{\rm diag} \, U_\ell^\dagger$, we
obtain the lepton mixing (PMNS) matrix in 
the weak charged lepton current 
\be 
\label{eq:pmns_def}
U = U_\ell^\dagger \, U_\nu~.
\ee
%
\\ 
From the analyzes of the currently existing neutrino oscillation 
data it was found \cite{thomas} that the present best-fit values 
of the CHOOZ and atmospheric neutrino mixing angles, 
$\theta_{13}$ and $\theta_{23}$, 
correspond to $|U_{e3}| = \sin\theta_{13} = 0$ and 
$\theta_{23} = \pi/4$, i.e., to
$|U_{\mu3}| = |U_{\tau 3}|$.  Accordingly,
the ``best-fit" PMNS matrix is given by
\be 
\label{eq:UBF}
U = 
\D
\left(
\bad  
\cos \theta_{12} & \sin \theta_{12} & 0 \\[0.3cm]
-\frac{\sin \theta_{12}}{\sqrt{2}} &  \frac{\cos \theta_{12}}{\sqrt{2}} 
&  -\frac{1}{\sqrt{2}} \\[0.3cm] 
-\frac{\sin \theta_{12}}{\sqrt{2}} & \frac{\cos \theta_{12}}{\sqrt{2}} 
&  \frac{1}{\sqrt{2}} \\[0.3cm]
\ea 
\right)\, , 
\ee
%
where we have not written the Majorana 
phases \cite{BHP80,SchValle80D81}
and used $\theta_{23} = -\pi/4$ in the usual Particle Data Group (PDG) 
parametrization 
of the PMNS matrix. One well-known 
possibility to construct this ``phenomenological'' 
mixing matrix  is to  require a $\mu$--$\tau$ exchange symmetry for 
the neutrino mass matrix in the basis in which the 
charged lepton mass matrix is diagonal \cite{mutau}. 
Well-known examples of neutrino 
mixing with $\mu$--$\tau$ symmetry are 
the bimaximal \cite{bima} and tri-bimaximal \cite{tri} mixing 
matrices
\be 
\label{eq:Usp}
U_{\rm bi} = \left(\bad  \frac{1}{\sqrt{2}} &  \frac{1}{\sqrt{2}} 
&  0 \\[0.3cm]
-\frac{1}{2} &  \frac{1}{2} &  -\frac{1}{\sqrt{2}} \\[0.3cm]
-\frac{1}{2} &  \frac{1}{2} &  \frac{1}{\sqrt{2}} \\[0.3cm]
\ea 
\right) \mbox{ and } 
U_{\rm tri} = 
\left(
\bad 
\sqrt{\frac{2}{3}} & \sqrt{\frac{1}{3}} & 0 \\[0.2cm]
-\sqrt{\frac{1}{6}} & \sqrt{\frac{1}{3}} & -\sqrt{\frac{1}{2}}  
\\[0.2cm]
-\sqrt{\frac{1}{6}} & \sqrt{\frac{1}{3}} & \sqrt{\frac{1}{2}}  
\ea 
\right)~.
\ee
%
A common feature of these two mixing matrices is $\theta_{23} = \pm \pi/4$ 
and $\theta_{13} = 0$, which is 
perfectly compatible with the current data. 
However, they differ in their prediction for the value of the 
solar neutrino mixing angle, namely, 
$\sin^2 \theta_{12} = 1/2$ and $1/3$, 
respectively. The best-fit value of 
$\sin^2 \theta_{12}$ determined from
the neutrino oscillation 
data is $\sin^2 \theta_{12} =0.30$.
Actually, $\sin^2 \theta_{12} = 1/2$ 
is ruled out by the data at more than $6\,\sigma$
\cite{BCGPRKL2}.  

A natural possibility to obtain a phenomenologically 
viable PMNS neutrino mixing
matrix, and to generate non-zero $|U_{e3}|$ and non-maximal 
$\theta_{23}$, is to assume that one of the two matrices in 
$U = U_\ell^\dagger \, U_\nu$ corresponds to 
Eq.~(\ref{eq:UBF}) or (\ref{eq:Usp}), 
and is ``perturbed'' by the second matrix  
leading to the required PMNS matrix. 
Following this assumption, corrections to
bimaximal \cite{bima_cor,FPR,alta,andrea,AK} and 
tri-bimaximal \cite{tri_cor,PlR,AK} mixing have previously 
been analyzed. 
For instance, scenarios in which the CKM quark mixing matrix corrects the 
bimaximal mixing pattern are important for models incorporating 
Quark-Lepton Complementarity 
(QLC) \cite{QLC0,QLC1,HR} (for earlier reference see \cite{SPAS94}).
Corrections to mixing scenarios 
with $\theta_{12} = \pi/4$ and $\theta_{13}=0$ 
were considered in \cite{PR} 
(motivated by the $L_e - L_\mu - L_\tau$ 
flavor symmetry \cite{lelmlt}) and in \cite{andrea}. 
The case with $\theta_{23} = \pi/4$ and 
$\theta_{13}=0$ has been 
investigated in Refs.~\cite{mutaulep0,mutaulep1,AK,Pi}. 

Up to now in most analyzes it 
has been assumed that $U_\nu$ possesses a form which leads 
to $\sin^2 \theta_{23} = 1/2$ 
and $\theta_{13}=0$. 
However, the alternative possibility of 
$\theta_{23} = \pi/4$ and $\theta_{13}=0$ originating from 
$U_\ell$ is phenomenologically equally viable. 
We are aware of only few 
papers in which that option is discussed 
\cite{alta,andrea,anki0,others}.
A detailed study is still lacking in the literature.
In the present article we perform, in particular, a
comprehensive analysis of this possibility.
We also revisit the case of 
$U_{e3} \neq 0$ and $\theta_{23} \neq \pi/4$
due to corrections from $U_\ell^\dagger$ and derive 
parametrization independent 
sum-rules for the relevant oscillation 
parameters. We point out 
certain ``subtleties'' in the identification 
of the relevant phases governing CP violation in neutrino 
oscillations with the Dirac phase of the standard 
parametrization of the PMNS matrix.\\

Our paper is organized as follows: Section~\ref{sec:form} briefly summarizes the formalism and the relevant matrices 
from which the neutrino mixing observables can be reconstructed. 
We analyze the possibility of $U_\nu$ leading to 
$\sin^2 \theta_{23} = 1/2$ and $\theta_{13}=0$ and 
being corrected by a non-trivial $U_\ell$ in 
Sec.~\ref{sec:unu}. 
In Sec.~\ref{sec:ulep} the alternative case of $U_\ell$ 
causing $\sin^2 \theta_{23} = 1/2$ and $\theta_{13}=0$
and being  modified  
by a non-trivial $U_\nu$ is discussed. 
Section \ref{sec:concl} contains our conclusions.


\section{\label{sec:form}Formalism and Definitions}

We will use the following parametrization of the PMNS matrix: 
\bea 
\label{eq:Upara}
U = V\, {\rm diag}(1,e^{i \alpha}, e^{i(\beta + \delta)}) 
= O_{23}(\theta_{23}) \, U_{13}(\theta_{13},\delta) \, 
O_{12}(\theta_{12}) \, 
{\rm diag}(1,e^{i \alpha}, e^{i(\beta + \delta)})
\\[0.3cm] 
= \left( 
\bad 
c_{12}  \, c_{13} & s_{12}  \, c_{13} & s_{13} \, e^{-i \delta}  \\[0.2cm] 
-s_{12}  \, c_{23} - c_{12}  \, s_{23}  \, s_{13}  \, e^{i \delta} 
& c_{12}  \, c_{23} -  \, s_{12}  \, s_{23}  \, s_{13}  \, e^{i \delta} 
& s_{23}  \, c_{13}  \\[0.2cm] 
s_{12}  \, s_{23} - c_{12}  \, c_{23}  \, s_{13}  \, e^{i \delta} & 
- c_{12}  \, s_{23} - s_{12}  \, c_{23}  \, s_{13}  \, e^{i \delta} 
& c_{23}  \, c_{13}  
\ea   
\right) 
{\rm diag}(1, e^{i \alpha}, e^{i (\beta + \delta)}) \, , 
\eea 
%
where 
$c_{ij} = \cos\theta_{ij}$, 
$s_{ij} = \sin\theta_{ij}$ and
$O_{ij}(\theta_{ij})$ is a $3\times 3$ orthogonal 
matrix of rotations on angle $\theta_{ij}$ in the 
$ij$-plane. We have also defined
\bea
\label{eq:U13}
U_{13}(\theta_{13},\delta) =  
\left(
\bad  
c_{13} & 0 & s_{13} \, e^{-i \delta}
\\[0.2cm] 
0 & 1 & 0
\\[0.2cm] 
- s_{13} \, e^{i \delta} & 0 & c_{13}
\ea 
\right)\,. 
\eea
%
Hereby we have included the Dirac CP violating
phase $\delta$ and the two Majorana CP violating 
phases $\alpha$ and $\beta$ \cite{BHP80,SchValle80D81}.
In general, all phases 
and mixing angles of $U$ are 
functions of the parameters characterizing
$U_\nu$ and $U_\ell$. 
It can be shown that \cite{PPR03,FPR} 
after eliminating the unphysical phases, $U$ can be written as 
$U = \tilde{U}_\ell^{\dagger} \, U_\nu$,
where in 
the most general case $U_\nu$ and $\tilde{U}_\ell$ 
are given by
\bea
\label{eq:Unu}
U_\nu = 
P \, \tilde{U}_\nu \, Q 
= {\rm diag}(1,e^{i \phi}, e^{i \omega}) \, 
\tilde{U}_\nu \, {\rm diag}(1,e^{i \sigma}, e^{i \tau}) \\[0.3cm] 
= P \, O_{23}(\theta^{\nu}_{23}) \, 
U_{13}(\theta^{\nu}_{13},\xi)\, O_{12}(\theta^{\nu}_{12}) \, Q
\\[0.3cm] 
= P \, 
\left(
\bad  
c_{12}^\nu \, c_{13}^\nu & s_{12}^\nu \, c_{13}^\nu & s_{13}^\nu 
\, e^{- i \xi}
\\[0.2cm] 
-s_{12}^\nu  \, c_{23}^\nu - c_{12}^\nu  \, s_{23}^\nu 
\, s_{13}^\nu  \, e^{i \xi} 
& c_{12}^\nu  \, c_{23}^\nu - s_{12}^\nu  \, s_{23}^\nu  
\, s_{13}^\nu  \, e^{i \xi} 
& s_{23}^\nu  \, c_{13}^\nu 
\\[0.2cm] 
s_{12}^\nu  \, s_{23}^\nu - c_{12}^\nu  \, c_{23}^\nu  
\, s_{13}^\nu  \, 
e^{i \xi} & 
- c_{12}^\nu  \, s_{23}^\nu - s_{12}^\nu  \, c_{23}^\nu  
\, s_{13}^\nu  \, e^{i \xi} 
& c_{23}^\nu  \, c_{13}^\nu 
\ea 
\right) \, Q ~,
\eea
where $P = {\rm diag}(1,e^{i \phi}, e^{i \omega})$, 
$Q = {\rm diag}(1,e^{i \sigma}, e^{i \tau})$ are rather important 
for the results to be obtained, and 
\bea
\label{eq:Ulep}
\tilde{U}_\ell = 
O_{23}(\theta^{\ell}_{23})\, 
U_{13}(\theta^{\ell}_{13},\psi)\, O_{12}(\theta^{\ell}_{12}) 
\\[0.3cm]
= \left(
\bad  
c_{12}^\ell \, c_{13}^\ell & s_{12}^\ell \, c_{13}^\ell 
& s_{13}^\ell \, e^{- i \psi}
\\[0.2cm] 
-s_{12}^\ell \, c_{23}^\ell - 
c_{12}^\ell \, s_{23}^\ell \, s_{13}^\ell \, 
e^{i \psi} 
& c_{12}^\ell \, c_{23}^\ell - s_{12}^\ell \, s_{23}^\ell \, 
s_{13}^\ell 
\, e^{i \psi} 
& s_{23}^\ell \, c_{13}^\ell   
\\[0.2cm] 
s_{12}^\ell \, s_{23}^\ell - c_{12}^\ell \, c_{23}^\ell \, 
s_{13}^\ell 
\, e^{i \psi} 
& - c_{12}^\ell \, s_{23}^\ell - s_{12}^\ell \, c_{23}^\ell \, 
s_{13}^\ell \, 
e^{i \psi} 
& c_{23}^\ell \, c_{13}^\ell 
\ea 
\right)~.
\eea
%
Here we have defined $c_{ij}^{\ell, \nu} 
= \cos \theta_{ij}^{\ell, \nu}$ and 
$s_{ij}^{\ell, \nu} 
= \sin \theta_{ij}^{\ell, \nu}$. 
Thus, $\tilde{U}_\nu$ and $\tilde{U}_\ell$ 
contain one physical CP violating phase each 
\footnote{In Section 
\ref{sec:ulep} it will be convenient to define instead 
of $U_\ell$ its transposed matrix as 
$U_\ell^T = O_{23}(\theta^{\ell}_{23})\, 
U_{13}(\theta^{\ell}_{13},\psi)\, O_{12}(\theta^{\ell}_{12})$. 
In addition, $U_\nu^\dagger = P \, \tilde{U}_\nu \, Q$ will 
be used there.}. 
The remaining four phases 
are located in the diagonal matrices  $P$ and $Q$. 
Note that $Q$ is ``Majorana-like'' \cite{FPR}, i.e., 
the phases $\sigma$ and $\tau$  
contribute only to the low energy 
observables related to the Majorana nature of 
the neutrinos with definite mass. 
Typically that are specific 
observables associated  
with $|\Delta L| = 2$ processes, like 
neutrinoless double beta decay 
$(A,Z) \rightarrow (A,Z+2) + e^- + e^-$ 
(see, e.g., \cite{STPNob04,BPP1}). 
In the following we will be interested  
in models and the phenomenological consequences that result if
$\tilde{U}_\nu$ corresponds to Eq.~(\ref{eq:UBF}), 
while $\tilde{U}_\ell$ contains 
comparatively small angles, and vice versa. 
It proves convenient to introduce the abbreviations
$\sin \theta_{ij}^{\ell, \nu} = \lambda_{ij} > 0 $ 
for the small quantities  
we will use as expansion parameters in our 
further analysis.

Turning to the observables, 
the sines of the three mixing angles 
of the PMNS matrix $U$ are given by 
\be
 \sin^2\theta_{13} =  |U_{e3}|^2~~,~~
\sin^2 \theta_{12} = \frac{\D |U_{e2}|^2}{\D 1 - |U_{e3}|^2}~~,~~
\sin^2 \theta_{23} =  \frac{\D |U_{\mu 3}|^2}{\D 1 - |U_{e3}|^2}~.
\label{sinij}
\ee
%
The expressions quoted above are in terms 
of the absolute values of the
elements of $U$, 
which emphasizes the independence of parametrization. 
 In the case of 3-$\nu$ mixing under discussion
there are, in principle, three independent 
CP violation rephasing invariants, associated with the 
three CP violating phases of the PMNS matrix.
The invariant related to the Dirac phase $\delta$ is given as
\be
J_{\rm CP} = {\rm Im}\left\{ 
U_{e1}^\ast \, U_{\mu 3}^\ast \, U_{e 3} \,
U_{\mu 1} \right\}~,
\label{JCP}
\ee
%
which controls the magnitude of
CP violation effects in neutrino oscillations and 
is a directly observable quantity~\cite{PKSP3nu88}. 
It is analogous to the 
rephasing invariant associated with the 
Dirac phase in the Cabibbo-Kobayashi-Maskawa
quark mixing matrix, introduced 
in Ref.~\cite{CJ85}. 
In addition to 
$J_{\rm CP}$, there are two rephasing 
invariants associated with the two Majorana 
phases in the PMNS matrix, 
which can be chosen as \footnote{The expressions 
for the invariants $S_{1,2}$ that we give here and will use further in the discussion 
 correspond to Majorana conditions 
for the fields of neutrinos with definite mass 
$\nu_j$ that do not contain phase factors, 
see, e.g., \cite{BPP1}.} \cite{JMaj87,ASBranco00} 
(see also \cite{BPP1}):
\be \D 
S_1 = {\rm Im}\left\{ U_{e 1} \, U_{e 3}^\ast \right\}\,,~~~
S_2 = {\rm Im}\left\{ U_{e2} \, U_{e 3}^\ast \right\}.
\ee
%
The rephasing invariants associated 
with the Majorana phases are 
not uniquely determined. Instead of $S_1$ 
defined above we could also have chosen
$S'_1 = {\rm Im}\{ U_{\tau 1}^\ast \, U_{\tau 2} \}$ 
or $S''_1 = {\rm Im}\{ U_{\mu 1} \, U_{\mu 2}^\ast \}$,  
while instead of $S_2$ we could have used 
$S'_2 = {\rm Im}\{ U_{\tau 2}^\ast \, U_{\tau 3} \}$ 
or $S''_2 = {\rm Im}\{ U_{\mu 2} \, U_{\mu 3}^\ast \}$.
The Majorana phases 
$\alpha$ and $\beta$, 
or $\beta$ and $(\beta - \alpha)$,
can be expressed in terms of the rephasing invariants in this way 
introduced \cite{BPP1}, for instance via 
$\cos \beta = 1 - S_1^2/|U_{e1} \, U_{e3}|^2$.
The expression for, e.g., $\cos \alpha$ in terms of 
$S'_1$ is somewhat more cumbersome (it involves also 
$J_{\rm CP}$) and we will not give it here.
Note that CP violation due to the Majorana phase 
$\beta$ requires that both
$S_1 = {\rm Im}\{ U_{e 1} \,U_{e 3}^\ast \}\neq 0$ and
${\rm Re}\{ U_{e 1} \,U_{e 3}^\ast \}\neq 0$.
Similarly, 
$S_2 = {\rm Im}\{ U_{e2}^\ast \,U_{e3} \} \neq 0$ 
would imply violation of the CP symmetry only 
if in addition ${\rm Re}\{ U_{e 2}^\ast \,U_{e 3} \} \neq 0$. 

Finally, let us quote the current data on 
the neutrino mixing angles
\cite{thomas,BCGPRKL2}: 
\begin{eqnarray}
\sss &=& 0.30^{+0.02, \, 0.10}_{-0.03, \, 0.06} ~,\nonumber\\
\sin^2\theta_{23} &=& 
0.50^{+0.08, \, 0.18}_{-0.07, \, 0.16} ~,\nonumber\\
|U_{e3}|^2  &=&0^{+0.012, \, 0.041}_{-0.000} ~,\nonumber
\end{eqnarray}
%
where we have given the best-fit values 
as well as the $1\,\sigma$ and $3\,\sigma$ allowed ranges. 

%
\section{\label{sec:unu}Maximal 
Atmospheric Neutrino Mixing and $U_{e3}=0$ 
from the Neutrino Mass Matrix}
%

In this Section we assume that maximal atmospheric neutrino 
mixing and vanishing $|U_{e3}|$ are realized in the limiting case, where 
$U_\ell$ corresponds to the unit matrix. 
We can obtain 
$\theta^{\nu}_{23} = -\pi/4$ and $\theta^{\nu}_{13} = 0$ 
by requiring $\mu$--$\tau$ exchange 
symmetry \cite{mutau,mutaulep1} of the neutrino 
mass matrix in the basis in that the 
charged lepton mass matrix is diagonal. 
Under this condition we have
\bea  \label{eq:mutau}
m_{\nu} = 
\left( 
\bad 
A_\nu & B_{\nu} & B_{\nu} \\[0.2cm]
\cdot & D_{\nu} + E_{\nu} & E_{\nu} - D_{\nu} 
\\[0.2cm] 
\cdot & \cdot & D_{\nu} + E_{\nu} 
\ea
\right)~, \\[0.3cm]  
\mbox{ with } 
A_{\nu} \equiv m_{1} \, c_{12}^2 + e^{-2 i \alpha} \, m_{2} 
\, s_{12}^2 ~~,~ 
B_{\nu} \equiv (e^{-2 i \alpha} \, m_{2}  - m_{1}) \, c_{12}  
\, s_{12} /\sqrt{2} ~~,~\\[0.3cm] 
D_{\nu} \equiv  e^{-2 i \beta} \, m_{3}/2 ~~,~  
E_{\nu} \equiv  \frac 12 \, 
(e^{-2 i \alpha} \, m_{2} \, c_{12}^2 + m_{1} \, s_{12}^2) 
~,
\eea
%
where $m_{1,2,3}$ are the neutrino masses. The indicated
symmetry is assumed to hold 
in the charged lepton mass basis, although the charged lepton 
masses are obviously not $\mu$--$\tau$ symmetric. 
However, such a scenario can, for example, be easily realized in models with different 
Higgs doublets generating 
the up- and down-like particle masses.

For the  sines of the ``small'' angles in the matrix $U_\ell$ 
we  introduce the convenient notation 
$\sin \theta_{ij}^\ell = \lambda_{ij} > 0$ with $ij = 12, 13, 23$. 
We obtain the following expressions for the observables relevant 
for neutrino oscillation in the case under consideration: 
\bea 
\label{eq:osc1}
\sin^2 \theta_{12} \simeq \sin^2 \theta_{12}^\nu - 
\frac{1}{\sqrt{2}} 
\, \sin 2 \theta_{12}^\nu \, 
\left( \lambda_{12} \, \cos \phi + \lambda_{13} \, 
\cos (\omega - \psi) 
\right) 
~,\\[0.2cm]
|U_{e3}| \simeq \frac{1}{\sqrt{2}} \, 
\left| 
\lambda_{12} \, e^{i \phi} - \lambda_{13} \, e^{i (\omega - \psi)} 
\right| 
~,\\[0.2cm]
\sin^2 \theta_{23} \simeq \frac 12 + \lambda_{23} 
\, \cos (\omega - \phi) - 
\frac 14 \left(\lambda_{12}^2  - \lambda_{13}^2 \right) 
+ \frac 12 \, \cos(\omega - \phi - \psi) \, 
\lambda_{12} \, \lambda_{13}
~,\\[0.2cm]
J_{\rm CP} \simeq \frac{1}{4 \sqrt{2}} \, \sin 2 \theta_{12}^\nu \, 
\left( 
\lambda_{12} \, \sin \phi - \lambda_{13} \, \sin (\omega - \psi) 
\right) 
~. 
\eea
%
Setting in 
these equations $\theta_{12}^\nu$ to $\pi/4$ 
(to $\sin^{-1} \sqrt{1/3}$) reproduces the formulas
from \cite{FPR} (also \cite{PlR}).\\

A comment on the CP phases is in order.
The relevant Dirac CP violating phase(s) can be identified 
from the expression for the rephasing invariant 
$J_{\rm CP}$: these are $\phi$ or $(\omega - \psi)$, depending 
on the relative magnitude of $\lambda_{12}$ and $\lambda_{13}$. 
However, within the approach we are employing, 
a Dirac CP violating phase appearing in 
$J_{\rm CP}$ does not necessarily coincide with the 
Dirac phase in the standard parametrization of the PMNS matrix. 
For illustration it is sufficient to consider 
the simple case of 
$\lambda_{12} \neq 0$ and 
$\lambda_{13} = \lambda_{23} = 0$. Working to leading order 
in $\lambda_{12}$, it is easy to find that in this 
case the PMNS matrix can be written as  
\bea
\label{eq:Ulam12}
U \simeq 
\tilde{P} \, 
\left(
\bad  
c_{12}^\nu \, e^{-i \phi} + 
\frac{\lambda_{12} \, s_{12}^\nu}{\sqrt{2}} 
& s_{12}^\nu \, e^{-i \phi} - 
\frac{\lambda_{12} \, c_{12}^\nu}{\sqrt{2}}
& \frac{\lambda_{12}}{\sqrt{2}}
\\[0.2cm] 
\lambda_{12}\,c_{12}^\nu \, e^{-i \phi} 
- \frac{s_{12}^\nu}{\sqrt{2}}
& \lambda_{12} \, s_{12}^\nu \, e^{-i \phi} + 
\frac{c_{12}^\nu}{\sqrt{2}}
& - \frac{1}{\sqrt{2}}  \\[0.2cm] 
- \frac{s_{12}^\nu}{\sqrt{2}} 
& \frac{c_{12}^\nu}{\sqrt{2}}
& \frac{1}{\sqrt{2}}
\ea 
\right) \, \tilde{Q}\,, 
\eea
%
where $\tilde{P} = {\rm diag}(e^{i \phi},e^{i \phi}, e^{i \omega})$ 
and $\tilde{Q} = {\rm diag}(1,e^{i \sigma}, e^{i \tau})$. 
The phase matrix $\tilde{P}$ can be eliminated from $U$ by 
a redefinition of the phases of the charged lepton fields. 
The Majorana phases $\alpha$ and 
$\beta' \equiv (\beta + \delta)$ 
can be directly identified (modulo $2\pi$) with $\sigma$ and $\tau$. 
It is clear from the expressions  
(\ref{eq:Upara}) and (\ref{eq:Ulam12}) 
for $U$, however, that the phase $\phi$ does not 
coincide with the Dirac phase $\delta$ of the 
standard parametrization of $U$.  
Actually, the phase $\phi$ could be directly identified 
with the Dirac CP violating phase of a different 
parametrization of the PMNS matrix, namely, 
the parametrization in which 
$\tilde{U}$ in Eq.~(\ref{eq:Ulam12})
is given by
\bea
\tilde{U} = O_{12}(\tilde{\theta}_{12})\, 
{\rm diag}(e^{- i \delta'}, 1, 1) \,
O_{23}(\tilde{\theta}_{23})\, O_{12}(\theta'_{12}) \, 
\label{eq:tU2} \\[0.3cm]
= 
\left( 
\bad
c_{12}' \, \tilde c_{12} \, e^{-i \delta'} 
- \tilde c_{23} \, s_{12}' \, \tilde s_{12} & 
\tilde c_{12} \, s_{12}' \, e^{-i \delta'} 
+ c_{12}' \, \tilde c_{23} \, \tilde s_{12} & 
\tilde s_{12} \, \tilde s_{23} \\[0.2cm]
-\tilde c_{12} \, \tilde c_{23} \, s_{12}' - 
c_{12}' \, \tilde s_{12} \, e^{-i\delta'} & 
c_{12}' \, \tilde c_{12} \, \tilde c_{23} - 
s_{12}' \, \tilde s_{12} \, e^{-i \delta'} & 
\tilde c_{12} \, \tilde s_{23} \\[0.2cm] 
s_{12}' \, \tilde s_{23} & 
-c_{12}' \, \tilde s_{23} & \tilde c_{23} 
\ea
\right)~. 
\eea
From this parametrization 
it would follow (using
$|U_{e3}| = \sin \tilde \theta_{23} \, \sin \tilde 
\theta_{12}$ and 
$|U_{\mu 3}/U_{\tau 3}|^2 = \cos^2 \tilde \theta_{12} 
\, \tan^2 \tilde \theta_{23}$) that 
$\tilde \theta_{12}$ should be small and 
that atmospheric neutrino mixing was governed in 
leading order by $\tilde \theta_{23}$. 
In the limit of $\tilde \theta_{23} = \pm\pi/4$ and $\tilde 
\theta_{12} = 0$ one would have
$|U_{e 2}/U_{e 1}|^2 = \tan^2 \theta_{12}'$. 
Hence, to leading order 
the solar neutrino mixing would be governed by 
$\theta_{12}'$ and leptonic CP violation in neutrino oscillations 
would be described by $J_{\rm CP} = 
-\frac 18 \, \sin 2 \theta_{12}' \, \sin 2 \tilde \theta_{12} 
\, \sin 2 \tilde \theta_{23} \, \sin \tilde \theta_{23} \, \sin \delta'$. 
We would recover Eq.~(\ref{eq:Ulam12}) from Eq.~(\ref{eq:tU2}) 
if we identified $\tilde{\theta}_{23} = - \pi/4$,
$\tilde{s}_{12} = - \lambda_{12}$,
$c'_{12} = c_{12}^\nu$,
$s'_{12} =  s_{12}^\nu$, and 
$\delta' = \phi$. 

 We are not going to use the parametrization 
(\ref{eq:tU2}) in the following. Instead, 
the three neutrino mixing angles $\theta_{13}$, 
$\theta_{12}$ and $\theta_{23}$ 
will be determined using the absolute values 
of the elements of the PMNS matrix, Eq.~(\ref{sinij}).
Concerning the issue of CP violation in 
neutrino oscillations, we will work only with the 
CP violating rephasing invariant $J_{\rm CP}$.
However, it is still useful to keep in mind that, 
as the example discussed above illustrates,
in the approach we are following the 
resulting Dirac CP violating phase, which 
is the source of CP violation 
in neutrino oscillations, cannot 
always be directly identified 
\footnote{The same 
conclusion is valid, e.g., for 
the Dirac phase in the relation given in Eqs.~(1) of  
the third and fourth articles quoted in Ref.~\cite{AK}.}  
with the Dirac CP violating phase 
of the standard parametrization 
(\ref{eq:Upara}) of the neutrino 
mixing matrix 
\footnote{The matrix $V$ in the parametrization 
(\ref{eq:Upara}) of the PMNS matrix, and the 
matrix $\tilde{U}$ in the parametrization 
(\ref{eq:tU2}) are connected by a unitary matrix: 
$V = W~\tilde{U}$. The latter reduces to the 
unit matrix (or to a diagonal phase matrix) 
only when the Dirac CP violating phases 
$\delta$ and $\delta'$, present in
$V$ and $\tilde{U}$, take CP conserving values:
$\delta = k\pi$, $\delta' = k'\pi$, $k,k'=0,1,2,\ldots$ 
In this case we can write $V = \tilde{U}$
and can express the angles of $V$ 
in terms of the angles of 
$\tilde{U}$, and vice versa.}.\\

Returning to Eq.~(\ref{eq:osc1}), we note that 
both $|U_{e3}|$ and $\sin^2 \theta_{23}$ 
do not depend on the mixing angle $\theta_{12}^\nu$.
The quantities $\lambda_{12}$ and $\lambda_{13}$ 
are crucial for the magnitudes of 
$|U_{e3}|$, $\sin^2 \theta_{12}$ and $J_{\rm CP}$, whereas they 
enter into the expression for $\sin^2 \theta_{23}$ 
only quadratically. In fact, 
$\sin^2 \theta_{23}$ receives first order 
corrections only from $\lambda_{23}$, which in turn 
contributes to the other 
observables only via terms proportional to $\lambda_{23}^3$.
Unless there are accidental cancellations,  
$|U_{e3}|$ is lifted from its zero value 
due to non-zero $\lambda_{12}$ and/or $\lambda_{13}$. Atmospheric neutrino 
mixing can be maximal, or very close to maximal, for instance if 
$\omega - \phi = \pi/2$. 
Note that 
$\lambda_{12}$ and $\lambda_{13}$ in the expressions for 
$\sin^2 \theta_{12}$, $|U_{e3}|$ and $J_{\rm CP}$
are multiplied by cosines and/or sines of the 
same phases $\phi$ and $(\omega - \psi)$, 
respectively.   This means that 
if the terms proportional to $\lambda_{12}$ (to $\lambda_{13}$) 
dominate over the terms proportional to 
$\lambda_{13}$ (to $\lambda_{12}$) -- we will refer 
to this possibility as 
$\lambda_{12}$($\lambda_{13}$)-dominance 
\footnote{More 
concretely, the conditions for, e.g.,  
$\lambda_{12}$-dominance are:
$|\lambda_{12} \, \cos\phi | \gg |\lambda_{13} \, 
\cos(\omega - \psi)|$ and  
$|\lambda_{12} \, \sin\phi| \gg |\lambda_{13} \, 
\sin(\omega - \psi)|$.} --  
we have \cite{FPR,AK,mutaulep1}:
\be \label{eq:sum0}
\sin^2 \theta_{12} = \sin^2 \theta_{12}^\nu 
- \sin 2 \theta_{12}^\nu \, 
|U_{e3}| \, \cos \gamma~,
\ee
%
where $\gamma = \phi~\,{\rm or}~(\psi - \omega)$ 
is the CP violating phase (combination) 
appearing in the expression for $J_{\rm CP}$, 
$J_{\rm CP} \propto \sin \gamma$. 
The relation (\ref{eq:sum0}) implies a correlation of 
the initial 12-mixing in $U_\nu$ with $|U_{e3}|$ and 
the observable CP violation in neutrino oscillations.
If $\tilde{U}_\nu$ is a bimaximal 
mixing matrix, we have $\sin^2 \theta_{12}^\nu = 1/2$
and $\cos \gamma$ has to take a value close to one 
(while $|U_{e3}|$ has to be relatively large) 
in order to obtain sufficiently non-maximal 
solar neutrino mixing. Consequently, 
in the case of $\lambda_{12}$($\lambda_{13}$)-dominance, 
CP violation would be suppressed 
even though $|U_{e3}|$ can be sizable. 
On the other hand, if 
$\tilde{U}_\nu$ is a
tri-bimaximal mixing matrix, 
we have $\sin^2 \theta_{12}^\nu = 1/3$ 
which already is in good 
agreement with the present data. 
Hence, $|U_{e3}|\cos \gamma$ has to 
be relatively small.
Consequently, CP violation can be sizable 
if $|U_{e3}|$ has a value close to the existing 
upper limit.
This interesting feature has first been 
noticed in Ref.~\cite{PlR}. 
Generally, in the case of 
$\lambda_{12}$($\lambda_{13}$)-dominance we 
get from Eq.~(\ref{eq:osc1}):
\be 
\label{eq:sum10}
\sin^2 \theta_{12} = \sin^2 \theta_{12}^\nu 
- 4\,J_{\rm CP}\,\cot\gamma\,.
\ee
where $\gamma = \phi$ 
($\gamma = \psi - \omega$)
for $\lambda_{12}$-dominance 
($\lambda_{13}$-dominance). 
%
The following ``sum-rule'' holds as well:
\be 
\label{eq:sum1}
\sin^2 \theta_{12} = \sin^2 \theta_{12}^\nu \pm 
\sqrt{|U_{e3}|^2 \, \sin^2 2 \theta_{12}^\nu - 16 \, J_{\rm CP}^2}~,
\ee
%
where the minus (plus) sign represents a positive (negative) 
cosine of the relevant Dirac CP violating phase. 
The sign ambiguity is unavoidable because the CP conserving quantity 
$\sin^2 \theta_{12}$ can only depend on the cosine of a 
CP violating phase, whereas 
any CP violating quantity like $J_{\rm CP}$ can only depend 
on the sine of this phase. Knowing the cosine of a 
phase will never tell us the sign of the sine 
\footnote{The same ambiguity will show up if 
one identifies the phase $\phi$ with the phase $\delta$ of a given 
parametrization of the PMNS matrix, as done, e.g., in Ref.~\cite{AK}. 
See also the comments given after Eq.~(\ref{eq:osc1}).}. 
Note that 
since all parameters in Eq.~(\ref{eq:sum1}) 
are rephasing invariant quantities, it  
can be applied to any parametrization of the 
PMNS matrix $U$ and of the matrix $\tilde{U}_\nu$. 
If $\tilde{U}_\nu$ is a
bimaximal (tri-bimaximal) mixing matrix, we get 
\be \label{eq:sum1sp}
\sin^2 \theta_{12} = 
\frac 12 \pm \sqrt{|U_{e3}|^2 
 - 16 \, J_{\rm CP}^2}~~\mbox{ and }~~ 
\sin^2 \theta_{12} = 
\frac 13  \left( 1 \pm 2 \sqrt{2} \, 
\sqrt{|U_{e3}|^2 - 6 \, J_{\rm CP}^2 }
\right)~,
\ee
%
respectively. The first relation has been obtained also in 
Ref.~\cite{HR}. Obviously, one has to choose here the negative 
sign. 

\begin{figure}
\begin{center}\vspace{-1cm}
\includegraphics[width=8cm,height=8cm]{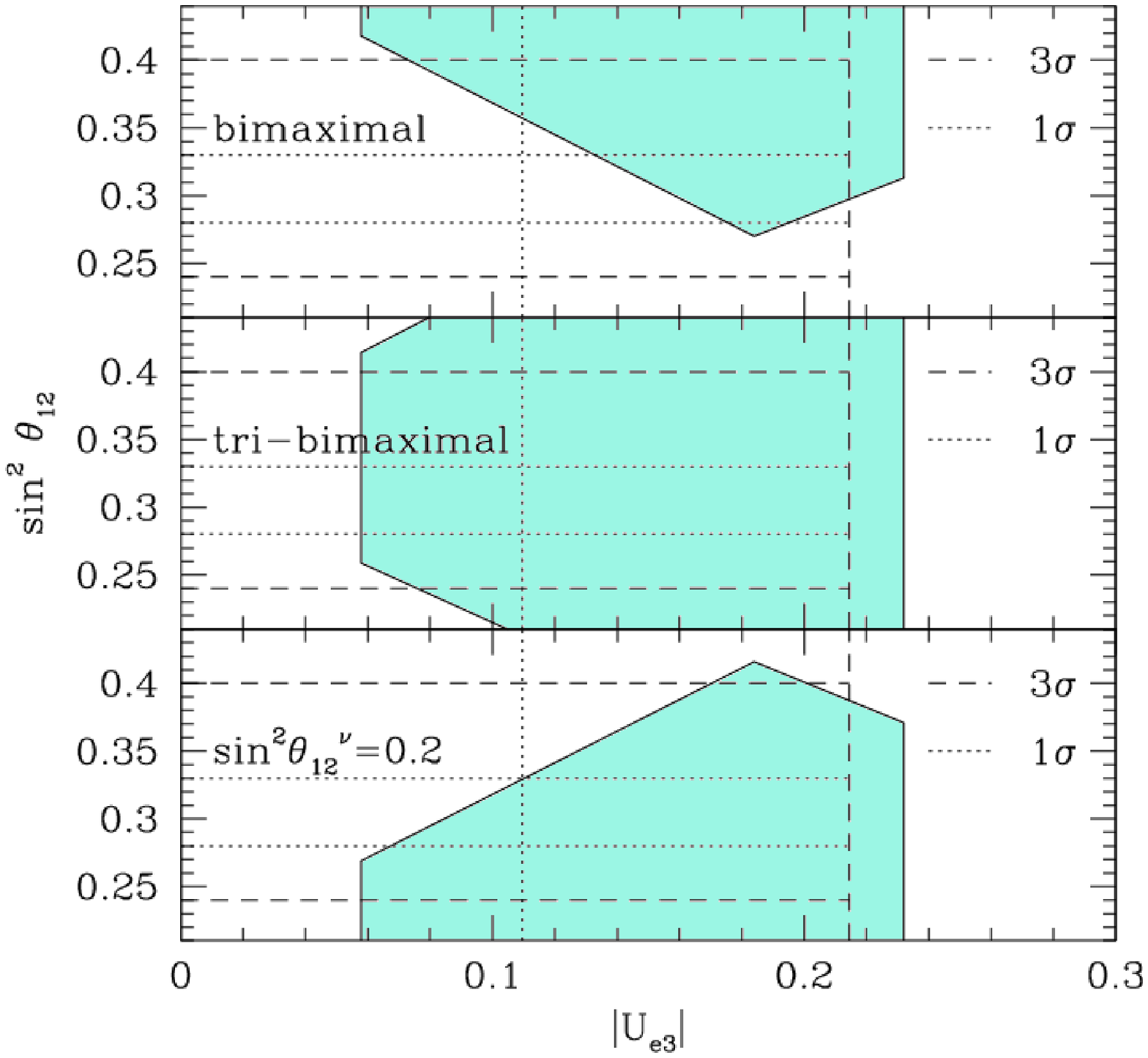}
\includegraphics[width=8cm,height=8cm]{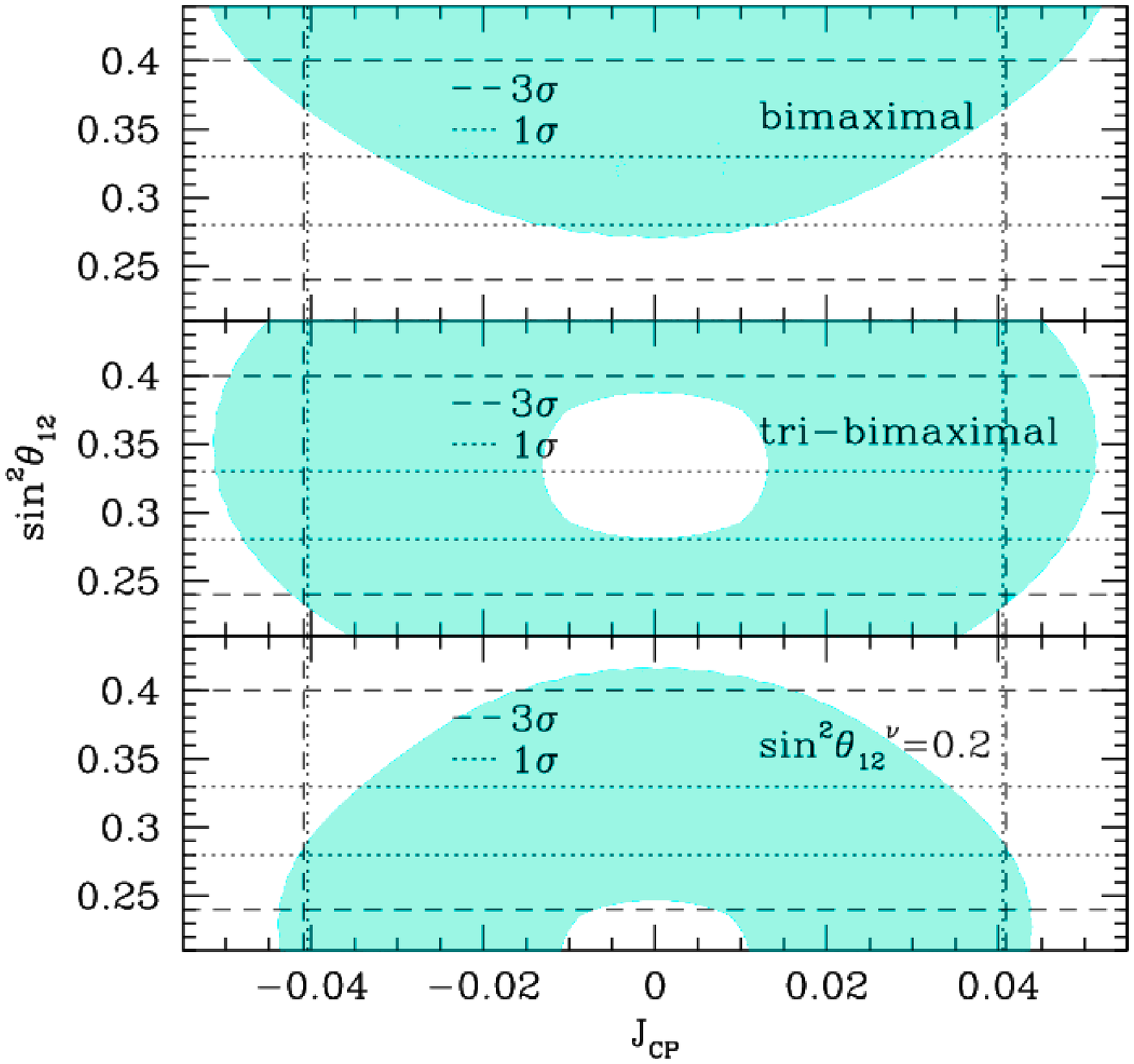}
\includegraphics[width=8cm,height=8cm]{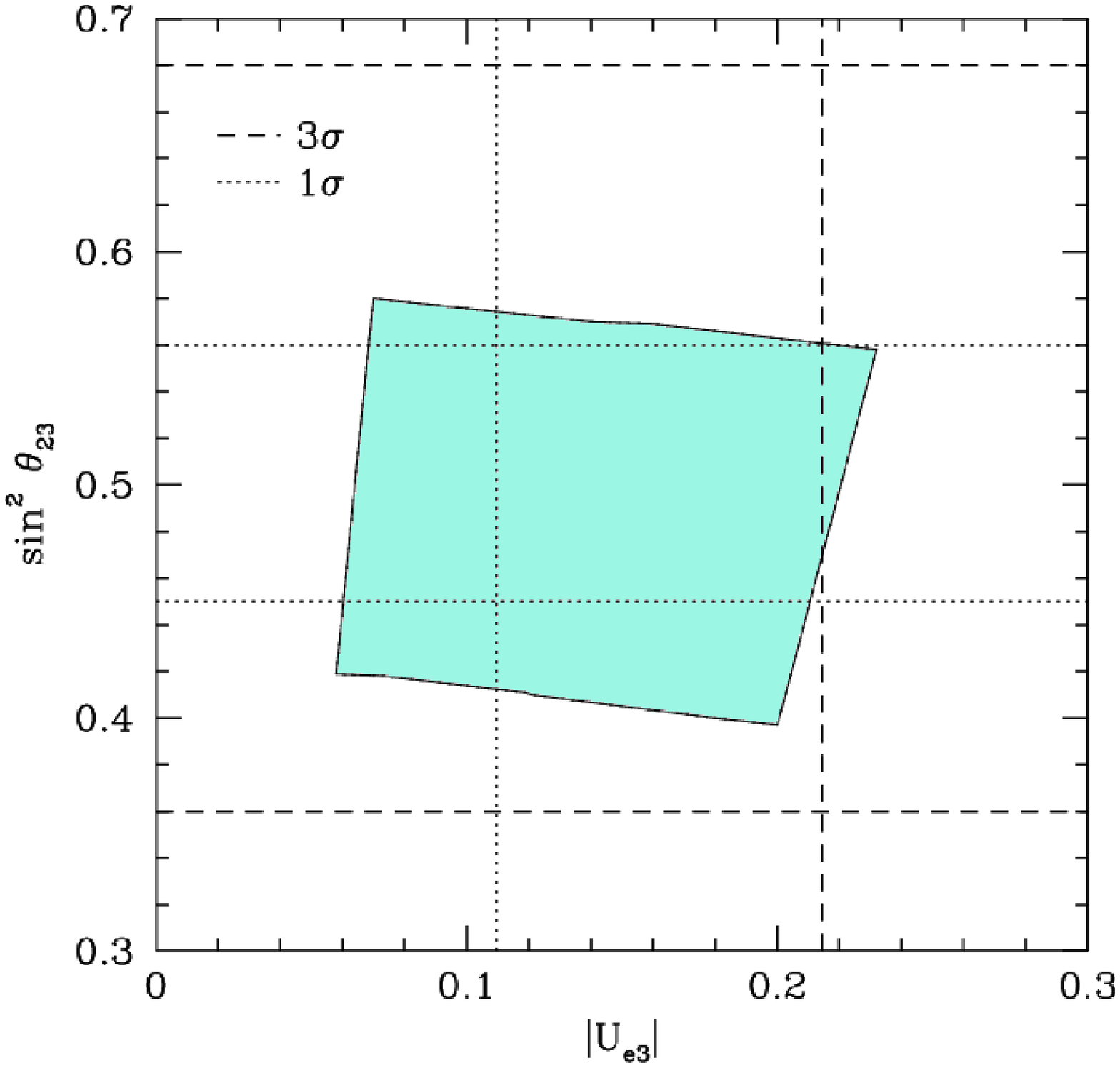}
\caption{\label{fig:obs1} Correlations 
resulting from $U = U_\ell^\dagger \, U_\nu$ 
if $U_\ell$ is CKM-like and $U_\nu$ has maximal 
$\theta_{23}^\nu$, vanishing $\theta_{13}^\nu$, 
but free $\theta_{12}^\nu$, for three representative values 
of $\theta_{12}^\nu$ (see text for details). The currently allowed 
$1\,\sigma$ and $3\,\sigma$ ranges of the observables are also indicated.}
\end{center}
\end{figure}

\begin{figure}
\begin{center}
\includegraphics[width=11cm,height=10cm]{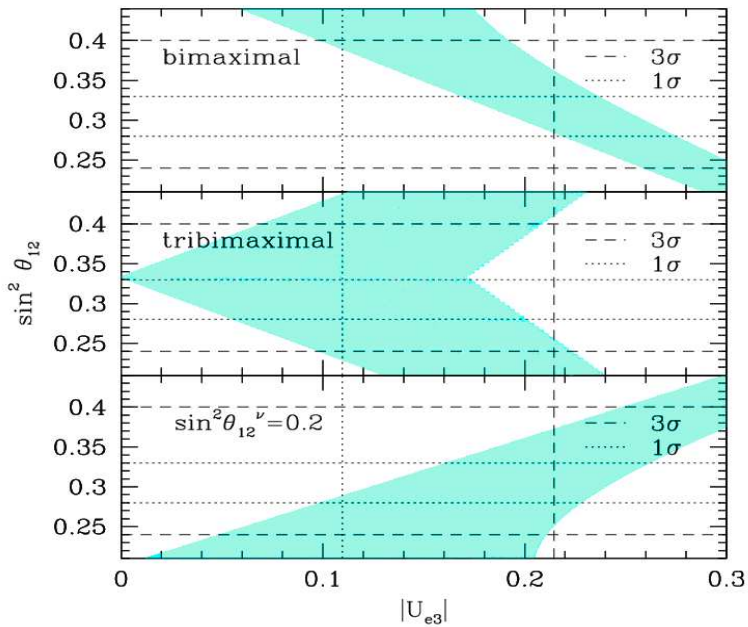}
\caption{\label{fig:obs1a}The sum-rule from Eq.~(\ref{eq:sum1}).}
\end{center}
\end{figure}

In Fig.~\ref{fig:obs1} we show the 
allowed parameter space for the 
exact equations in the cases of $\sin^2\theta^\nu_{12}=1/2$ 
(bimaximal mixing), $1/3$ (tri-bimaximal mixing) and $0.2$. 
We have chosen the 
$\lambda_{ij}$ to obey a CKM-like hierarchy: 
$0.1 \le \lambda_{12} \le 0.3$, 
$0.02 \le \lambda_{23} \le 0.08$ and $0 \le \lambda_{13} \le 0.01$.
As $|U_{e3}|$ and $\sin^2 \theta_{23}$ are independent of 
$\theta_{12}^\nu$ we have plotted these observables only once. 
The chosen ranges of the $\lambda_{ij}$ lead from 
Eq.~(\ref{eq:osc1}) to a lower limit of 
$|U_{e3}| \gs 0.09/\sqrt{2} \simeq 0.06$, 
as is seen in the figure.
Improved future limits on the range of 
$\sin^2 \theta_{12}$ and, in particular, on the 
magnitude of $|U_{e3}|$ can give us 
valuable information on the structure of $U_\ell$. 
The allowed parameter space of $\sin^2 \theta_{23}$ 
is roughly half of its allowed $3\,\sigma$ range. 
The interplay of $\theta_{12}^\nu$ and leptonic CP violation 
in neutrino oscillations mentioned above results in the 
``falling donut'' structure when $J_{\rm CP}$ is plotted 
against $\sin^2 \theta_{12}$. 
We can also directly plot the sum-rule from
Eq.~(\ref{eq:sum1}), which is shown in Fig.~\ref{fig:obs1a}. 
As a consequence of varying the observables in Eq.~(\ref{eq:sum1})
we can extend the parameter space to smaller values of $|U_{e3}|$.
In fact, if $U_\nu$ corresponds to tri-bimaximal mixing,
$U_{e3}$ is allowed to vanish. 
Equation~(\ref{eq:osc1}) can be used to understand
the results in Fig.~\ref{fig:obs1a}: if, for instance, we
have $\sin^2 \theta_{12}^\nu = 1/2$, 
the experimental upper limit of $(\sin^2 \theta_{12})_{\rm max} = 0.4$
implies that $|U_{e3}| \ge 1/2 - (\sin^2 \theta_{12})_{\rm max}
\simeq 0.1$. 
On the other hand, for $\sin^2 \theta_{12}^\nu = 0.2$,
and therefore $\sin 2 \theta_{12}^\nu = 0.8$, we have with
$(\sin^2 \theta_{12})_{\rm min} = 0.24$ that
$|U_{e3}| \ge ((\sin^2 \theta_{12})_{\rm min} - 0.20)/0.8 \simeq 0.05$,
which is in agreement with the figure. 
A more stringent limit on, or a value of, 
$|U_{e3}|^2 \ls 0.01$ would strongly disfavor (or rule out) 
the simple bimaximal mixing scenario.

The equations given up to this point 
are also valid if neutrinos are Dirac particles. 
We will discuss now briefly the observables 
describing the CP violation associated with the Majorana 
nature of the massive neutrinos. We find that in the case 
under discussion
\bea \label{eq:S1S21}
S_1 \simeq -\frac{1}{ \sqrt{2}} \, \cos \theta_{12}^\nu \, 
\left(\lambda_{12} \, \sin (\phi + \tau) - \lambda_{13} \, 
\sin (\omega - \psi + \tau) 
\right)
~,\\[0.2cm]
S_2 \simeq \frac{ 1}{\sqrt{2}} \, \sin \theta_{12}^\nu \, 
\left(\lambda_{12} \, \sin (\sigma - \phi - \tau) - \lambda_{13} \, 
\sin (\sigma - (\omega - \psi  + \tau)) 
\right) 
~.
\eea
According to the 
parameterization of Eq.~(\ref{eq:Upara}), we have
$S_1 = -c_{12} \, c_{13} \, s_{13} \, \sin \beta$ and 
$S_2 = s_{12} \, c_{13} \, s_{13} \, \sin (\alpha - \beta)$. 
Hence, we find that in the case of $\lambda_{12}$-dominance, 
$\beta$ is associated 
\footnote{Actually such an identification 
is always valid modulo $2\pi$. For simplicity, we will omit stating 
this explicitly from here on.} with $ \phi + \tau$, 
while if the terms proportional to $\lambda_{13}$ 
dominate over the terms  proportional to $\lambda_{12}$,  
the phase $\beta$ is associated with $\psi - \omega - \tau$. 
In both cases $\alpha$ is associated with $\sigma$.  

Obviously, if $\sigma = 0$ we get in the case of 
$\lambda_{12}$- or $\lambda_{13}$-dominance that 
$S_1 \simeq S_2 \tan \theta^{\nu}_{12}$.
We note also that, as it follows from Eqs.~(\ref{eq:osc1}) 
and (\ref{eq:S1S21}), 
for $\tau \simeq 0$ the Dirac CP violating phase $\delta$ 
will coincide with the Majorana CP violating phase $\beta$.\\

The most natural possibility for the structure of 
$U_\ell$ is that it is 
``CKM-like'', i.e., $\lambda_{23} = A \, \lambda_{12}^2$ and 
 $\lambda_{13} = B \, \lambda_{12}^3$ with $A$ and $B$ of order one. 
The resulting equations are
\bea \label{eq:osc1a}
\sin^2 \theta_{12} \simeq 
\sin^2 \theta_{12}^\nu - \frac{1}{\sqrt{2}} \, \cos \phi \, 
\sin 2 \theta_{12}^\nu \, \lambda_{12} + 
\frac{1}{2} \,  \cos 2 \theta_{12}^\nu \, \lambda_{12}^2
~,\\[0.2cm]
|U_{e3}| \simeq \frac{\D \lambda_{12}}{\D \sqrt{2}}~,\\[0.2cm]
\sin^2 \theta_{23} \simeq \frac 12 - \frac 14 \, 
\left( 
1 - 4 \, B \, \cos (\omega - \phi) 
\right) \, \lambda_{12}^2~,\\[0.2cm]
J_{\rm CP} \simeq \frac{1}{4\sqrt{2}} \, \lambda_{12}  
\sin 2 \theta_{12}^\nu \,\sin \phi \,, ~
\eea
%
plus cubic terms. The sum-rule in Eq.~(\ref{eq:sum1}) is of course 
valid. For the invariants describing the Majorana phases 
we have 
\bea
S_1 \simeq -\frac{1}{ \sqrt{2}} \, \left( 
\lambda_{12}\, \cos \theta_{12}^\nu \, \sin (\phi + \tau)   + 
 \lambda_{12}^2 \, \sin \theta_{12}^\nu \, \sin\tau
\right)~,\\[0.2cm]
S_2 \simeq -\frac{1}{ \sqrt{2}} \, \left( 
\lambda_{12} \, \sin \theta_{12}^\nu \, \sin (\phi - \sigma + \tau)  + 
\lambda_{12}^2 \, \cos \theta_{12}^\nu \, \sin (\sigma - \tau)   
\right)~.
\eea

\section{\label{sec:ulep}Maximal 
Atmospheric Mixing and $U_{e3}=0$ from the 
Charged Lepton Mass Matrix}
Now we study the equally interesting possibility that maximal 
$\theta_{23}$ and vanishing $|U_{e3}|$ are realized in the limiting case, where $U_\nu$ is equivalent to the unit matrix. 
In this scenario we have
\be \label{eq:Uell2}
U_\ell^\dagger = U_\ell^T = 
\D
\left(
\bad  
c_{12}^\ell &  s_{12}^\ell & 0 \\[0.3cm] \D
-\frac{s_{12}^\ell}{\sqrt{2}} & \D \frac{c_{12}^\ell}{\sqrt{2}} 
&  \D -\frac{1}{\sqrt{2}} \\[0.3cm] 
\D -\frac{s_{12}^\ell}{\sqrt{2}} & \D \frac{c_{12}^\ell}{\sqrt{2}} 
& \D \frac{1}{\sqrt{2}} \\[0.3cm]
\ea 
\right)\, ,
\ee
%
where we have to define $U_\ell^T =  O_{23}(\theta^{\ell}_{23})\, 
U_{13}(\theta^{\ell}_{13},\psi)\, O_{12}(\theta^{\ell}_{12})$ 
in order to have the rotations in the correct order, 
cf.~Eq.~(\ref{eq:Ulep}). 
Note that $U_\ell$ is real and therefore 
$m_\ell^\dagger \, m_\ell = U_\ell \, (m_\ell^{\rm diag})^2 \, 
U_\ell^\dagger$ is symmetric. Reconstructing this matrix gives 
\bea \label{eq:mlep} 
m_\ell^\dagger \, m_\ell = \\[0.2cm]
\hspace{-.2cm}\left( 
\bad 
m_e^2 \, (c_{12}^\ell)^2 +  \frac 12 \, (s_{12}^\ell)^2 
\, (m_\mu^2 + m_\tau^2) 
& c_{12}^\ell \, s_{12}^\ell \, 
(m_e^2 - \frac 12 \, m_\mu^2 - \frac 12 \, m_\tau^2) 
&  \frac 12 \, s_{12}^\ell \, (m_\mu^2 - m_\tau^2) \\[0.3cm]
\cdot & m_e^2 \, (s_{12}^\ell)^2 +  \frac 12 \, (c_{12}^\ell)^2 
\, (m_\mu^2 + m_\tau^2) & 
 \frac 12 \, c_{12}^\ell \, (m_\mu^2 - m_\tau^2) \\[0.3cm]
\cdot & \cdot & \frac 12 \, (m_\mu^2 + m_\tau^2)
\ea
\right)\; ,
\eea
%
which does not obey a simple exchange symmetry as 
the neutrino mass matrix in 
Eq.~(\ref{eq:mutau}). However, there are relations 
between the entries: for instance, if we additionally
assume $\theta_{12}^\ell = - \pi/4$, we find 
\bea  \label{eq:mleptau}
 m_\ell^\dagger \, m_\ell = 
\left( 
\bad 
A_\ell  + D_\ell & A_\ell - D_\ell & B_\ell \\[0.2cm]
\cdot & A_\ell + D_\ell  & B_\ell \\[0.2cm] 
\cdot & \cdot & 2 \, A_\ell 
\ea
\right)~, \\[0.3cm]  
\mbox{ with } 
A_\ell \equiv \frac 14 \, (m_\mu^2 + m_\tau^2) ~~,~ 
B_\ell \equiv (m_\tau^2 - m_\mu^2)/\sqrt{8} ~~,~
D_\ell \equiv \frac 12 \, m_e^2~.
\eea
%
Discrete symmetries might be capable of generating such a texture. 
Another hint towards a possible origin of such a matrix can be 
obtained by noting that due to $m_\tau^2 \gg m_\mu^2 \gg m_e^2$ the 
entries are all of similar magnitude \cite{alta}, 
and therefore $m_\ell^\dagger \, m_\ell$ 
resembles the mass matrices of the 
``flavor democratic'' type.\\

We have to multiply $U_\ell^\dagger = U_\ell^T$ 
from Eq.~(\ref{eq:Uell2}) with the matrix $U_\nu$ to obtain the 
PMNS matrix. Let us first assume that $U_\nu$ is given by the 
hermitian adjoint of Eq.~(\ref{eq:Unu}): 
$U_\nu^\dagger = P \, O_{23}(\theta^{\nu}_{23}) \, 
U_{13}(\theta^{\nu}_{13},\xi)\, O_{12}(\theta^{\nu}_{12}) \, Q$. 
This will bring the 
12-rotations of $U_\ell$ and $U_\nu$ directly together and, in 
absence of phases, would lead to
$\theta_{12} = \theta^{\ell}_{12} - \theta^{\nu}_{12}$,  
a feature which makes this 
possibility interesting for Quark-Lepton 
Complementarity scenarios \cite{QLC0,QLC1,HR}. 
For the neutrino oscillation observables we get 
\bea
\label{eq:osc2a}
\sin^2 \theta_{12} \simeq \sin^2 \theta_{12}^\ell - 
\lambda_{12} \, \sin 2 \theta_{12}^\ell \, \cos \sigma + 
\frac 14 \, \left(\lambda_{13}^2 - \lambda_{23}^2 \right)\, 
\sin^2 2 \theta_{12}^\ell 
+ \lambda_{12}^2 \, \cos 2 \theta_{12}^\ell\,  
~,\\[0.2cm]
|U_{e3}| \simeq \left| \lambda_{23} \, \sin \theta_{12}^\ell  
 + \lambda_{13} \, \cos \theta_{12}^\ell \, e^{i(\xi - \sigma)} \right|
~,\\[0.2cm]
\sin^2 \theta_{23} \simeq 
\frac 12 \, + \, \lambda_{23} \, \cos \theta_{12}^\ell \, 
\cos(\xi - \sigma + \tau)  
 - \lambda_{13} \, \sin \theta_{12}^\ell \, \cos \tau  
~,\\[0.2cm]
J_{\rm CP} \simeq 
-\frac 14 \, \sin 2 \theta_{12}^\ell \, \left(  \lambda_{23} \,
\sin \theta_{12}^\ell \, \sin(\xi - \sigma + \tau) 
+ \lambda_{13} \, \cos \theta_{12}^\ell \, \sin \tau  
\right)
~.
\eea 
The parameter $\lambda_{12}$ is crucial for obtaining a
sufficiently non-maximal angle $\theta_{12}$ in the case of a
bimaximal $U_\ell^\dagger$. However, $\lambda_{12}$ 
appears only in terms proportional to $\lambda_{12}^3$ in 
$|U_{e3}|$, $\sin^2 \theta_{23}$ and $J_{\rm CP}$.  
In these latter observables $\lambda_{13}$ and $\lambda_{23}$ 
are multiplied by the sines or cosines of the same phases. 
As a consequence, we can write down a correlation analogous 
to the one given in Eq.~(\ref{eq:sum10}). Namely, if 
the terms  proportional to $\lambda_{23}$ dominate over the
terms proportional to $\lambda_{13}$ 
(``$\lambda_{23}$-dominance''), we have 
\be 
\label{eq:sum20}
\sin^2 \theta_{23} \simeq \frac 12 - 2 \, J_{\rm CP} \, 
\frac{\cot (\xi - \sigma + \tau)}{\sin^2 \theta_{12}^\ell}~.
\ee
The analogue of the sum-rule in Eq.~(\ref{eq:sum1}) is
\be 
\label{eq:sum2}
\sin^2 \theta_{23} \simeq \frac 12 \pm 
\frac{1}{\sin^2 \theta_{12}^\ell} \, \sqrt{|U_{e3}|^2 \, 
\cos^2 \theta_{12}^\ell \, \sin^2 \theta_{12}^\ell - 4 \, 
J_{\rm CP}^2}
\,,
\ee
%
where the plus (minus) sign corresponds to 
$\cos (\xi - \sigma + \tau) > 0$ ($\cos (\xi - \sigma + \tau) < 0$). 
In this scenario the value of the 
atmospheric neutrino mixing angle 
is correlated with the magnitude of CP violation 
effects in neutrino oscillations. 
In the case of $\sin^2 \theta_{12}^\ell = 1/2$ or $1/3$ we find 
\be 
\label{eq:sum2sp}
\sin^2 \theta_{23} - \frac 12 = \pm 
\sqrt{|U_{e3}|^2 - 16 \, J_{\rm CP}^2} \mbox{ or } 
\sin^2 \theta_{23} - \frac 12 = 
\pm \sqrt{2} \, \sqrt{|U_{e3}|^2 - 18 \, J_{\rm CP}^2}~,
\ee
%
respectively. The first relation has been obtained also in 
Ref.~\cite{HR}.  A high precision measurement of 
$\sin^2 \theta_{23}$, combined with 
a sufficiently stringent limit on, or a relatively 
small measured value of, $|U_{e3}|^2$ might allow 
to discriminate between the simple bimaximal and tri-bimaximal 
mixing scenarios we are considering.

\begin{figure}
\begin{center}\vspace{-1cm}
\includegraphics[width=8cm,height=8cm]{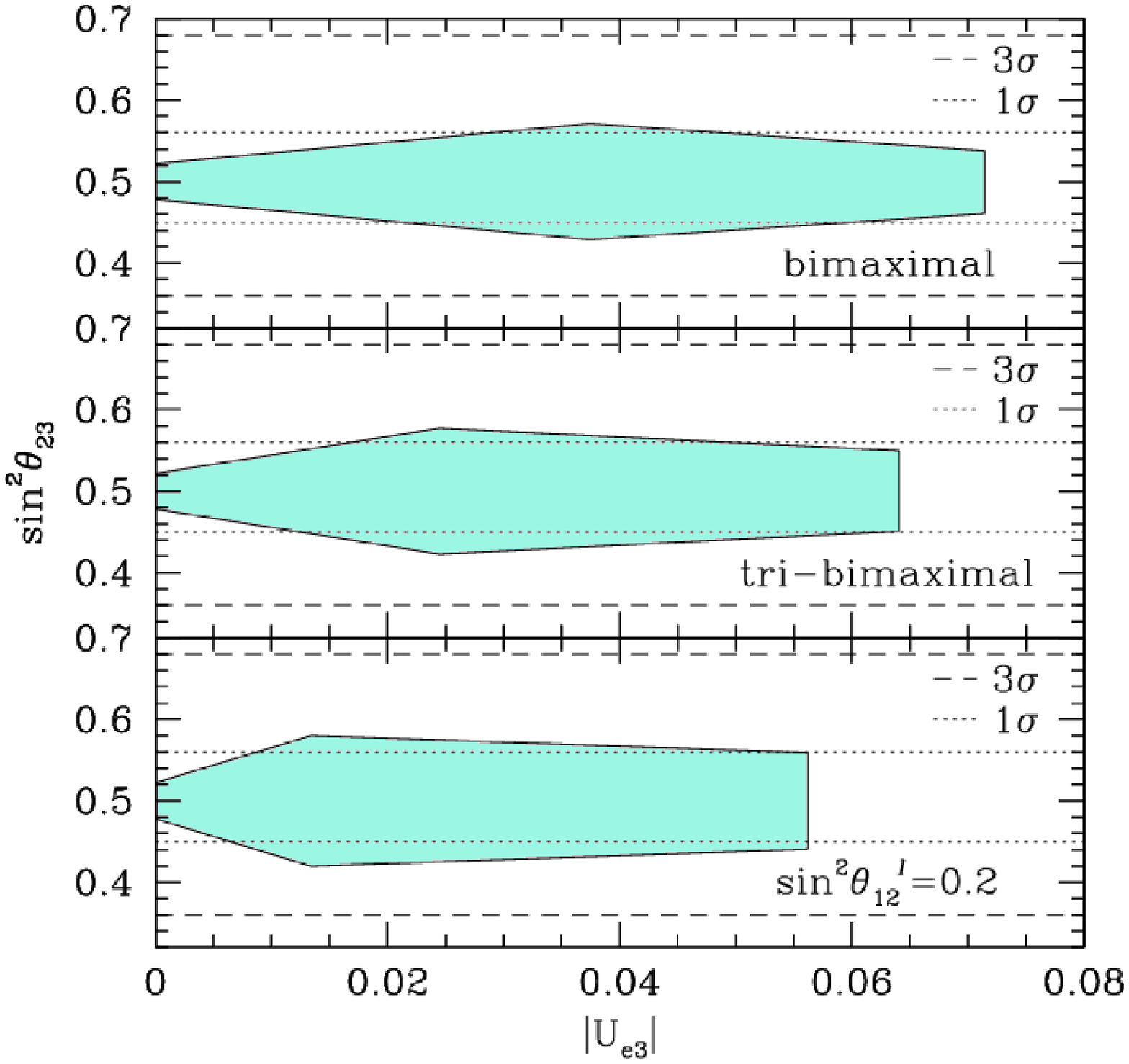}
\includegraphics[width=8cm,height=8cm]{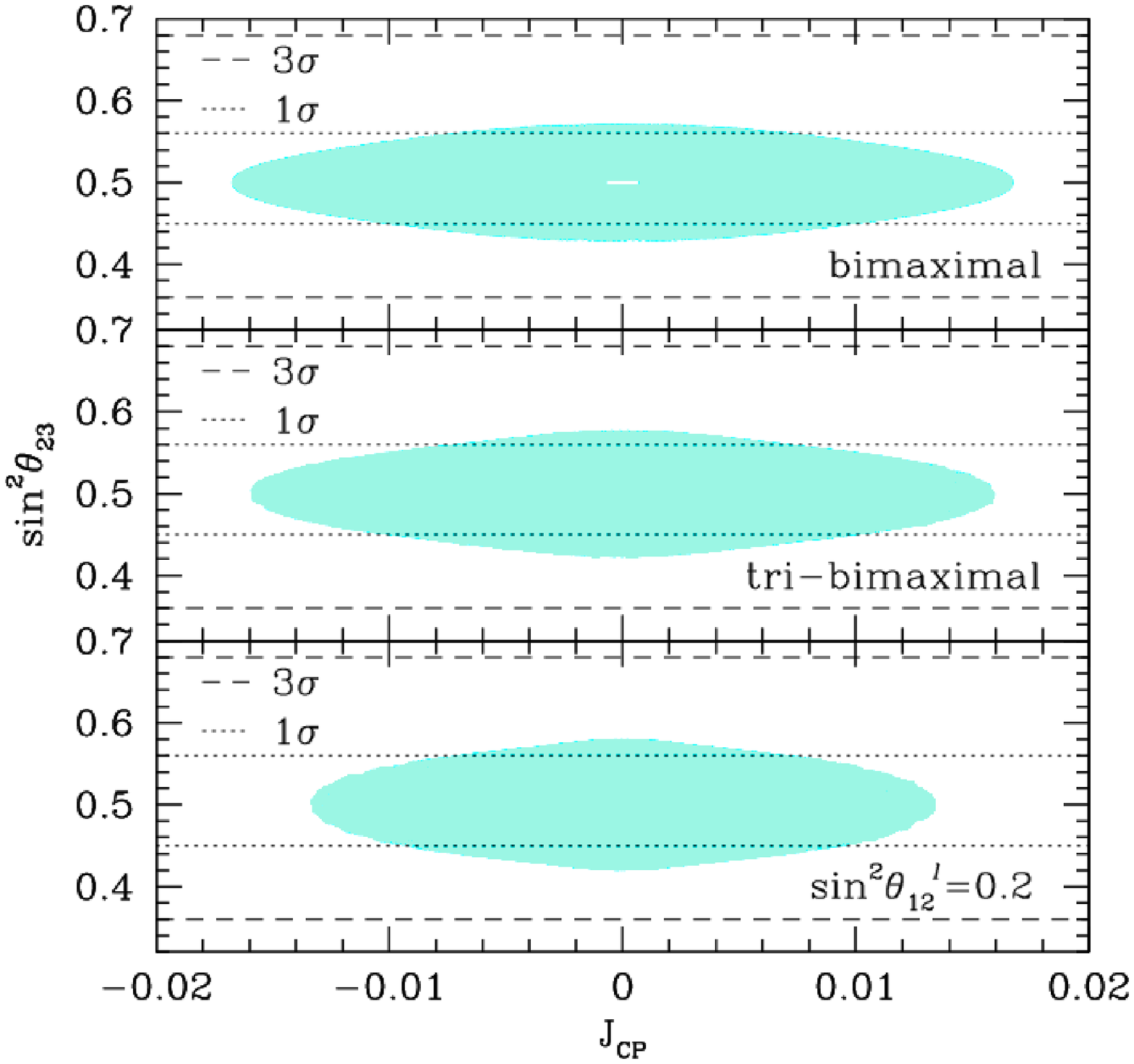}
\includegraphics[width=8cm,height=8cm]{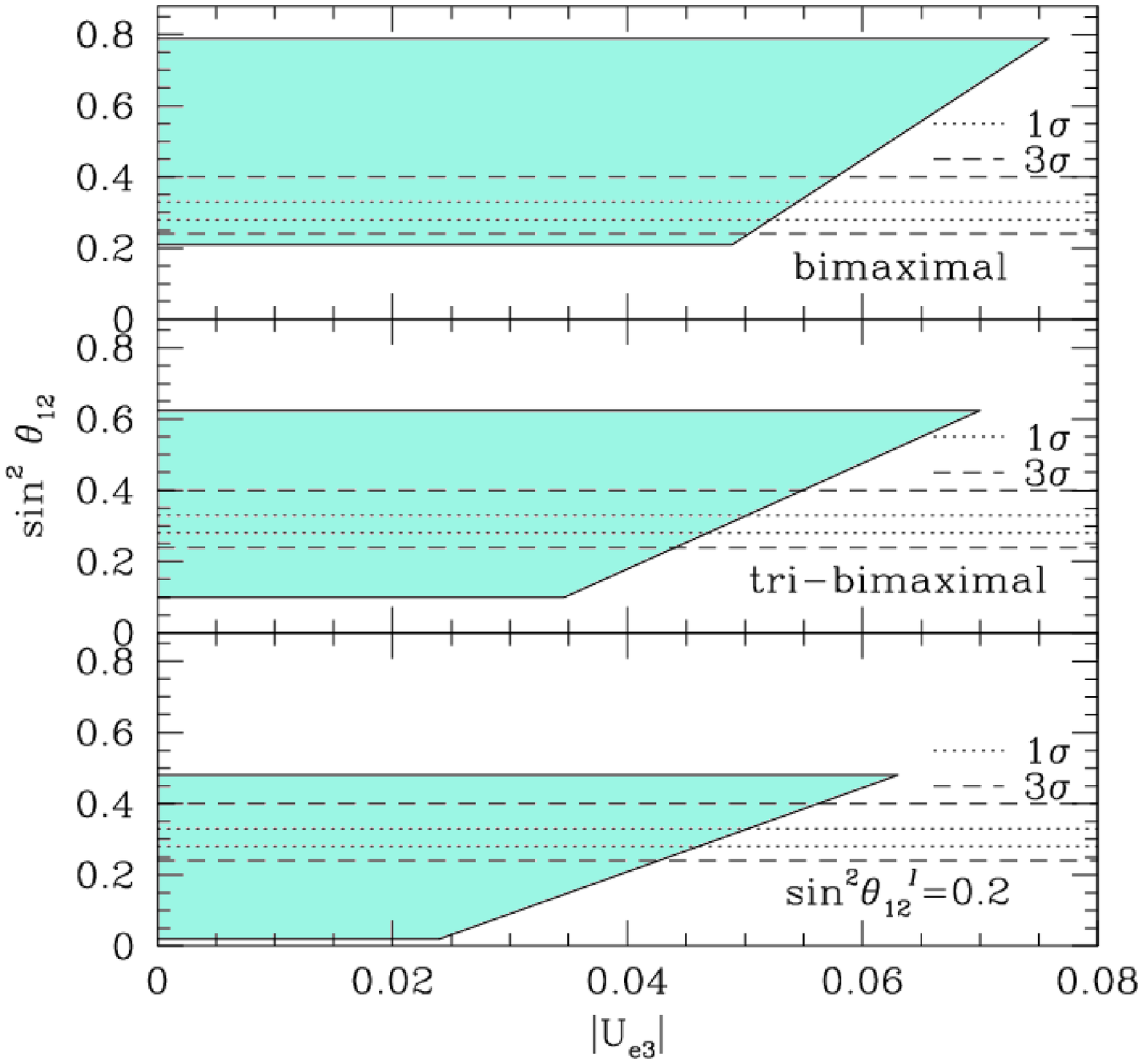}
\caption{\label{fig:obs2} Correlations 
resulting from $U = U_\ell^\dagger \, U_\nu$ 
if $U_\nu$ is CKM-like and $U_\ell^\dagger$ has maximal $\theta^{\ell}_{23}$ 
and vanishing $\theta^{\ell}_{13}$, 
but free $\theta_{12}^\ell$. The results shown 
correspond to three representative values of $\theta_{12}^\ell$. 
The currently allowed 
$1\,\sigma$ and $3\,\sigma$ ranges of the observables are also indicated.}
\end{center}
\end{figure}

The corresponding relations in the case 
of $\lambda_{13}$-dominance are 
\be 
\label{eq:sum30}
\sin^2 \theta_{23} \simeq \frac 12 + 2 \, J_{\rm CP} \, 
\frac{\cot \tau}{\cos^2 \theta_{12}^\ell}~,
\ee
and
\be 
\label{eq:sum3}
\sin^2 \theta_{23} \simeq \frac 12 \mp 
\frac{1}{\cos^2 \theta_{12}^\ell} \, \sqrt{|U_{e3}|^2 \, 
\cos^2 \theta_{12}^\ell \, \sin^2 \theta_{12}^\ell - 4 \, J_{\rm CP}^2}\,,
\ee
where the minus (plus) sign corresponds to 
$\cos \tau > 0$ ($\cos \tau < 0$).
The results for 
$\sin^2 \theta_{12}^\ell = 1/2$ or $1/3$ 
can be easily obtained as 
\be 
\label{eq:sum3sp}
\sin^2 \theta_{23} - \frac 12 = \mp 
\sqrt{|U_{e3}|^2 - 16 \, J_{\rm CP}^2} \mbox{ or } 
\sin^2 \theta_{23} - \frac 12 = 
\mp \frac{1}{\sqrt{2}} \, \sqrt{|U_{e3}|^2 - 18 \, J_{\rm CP}^2}~.
\ee
%
In Fig.~\ref{fig:obs2} we show the allowed parameter space for the 
exact equations in the cases of $\sin^2\theta^\ell_{12}=1/2$ 
(bimaximal), 
$1/3$ (tri-bimaximal) and $0.2$. 
We have chosen again the 
$\lambda_{ij}$ to follow a CKM-like hierarchy with 
$0.1 \le \lambda_{12} \le 0.3$, 
$0.02 \le \lambda_{23} \le 0.08$ and $0 \le \lambda_{13} \le 0.01$. 
Note that -- in contrast to the first scenario -- 
$|U_{e3}|$ is much smaller and 
can even vanish exactly not only 
when $\sin^2\theta^\ell_{12}=1/3$, but also 
for $\sin^2\theta^\ell_{12}=1/2~{\rm or}~0.2$.
Moreover, the range of the 
$\lambda_{ij}$ and the dependence of $\sin^2 \theta_{23}$ on them  
lead to the absence of a characteristic donut-like 
structure as seen in Fig.~\ref{fig:obs1}. 
For a CKM-like $U_\nu$, the importance 
of $\sin^2 \theta_{12}^\ell$ for $\sin^2 \theta_{23}$ and 
$|U_{e3}|$ is not as strong as it is the 
first scenario considered in Sec.~\ref{sec:unu}. 
As mentioned above, the value of $\sin^2 \theta_{12}^\ell$ 
is important mainly for the required magnitude of 
$\lambda_{12}$ which 
is responsible only for subleading contributions to 
the other parameters. As in the first scenario, atmospheric neutrino 
mixing can be maximal. If $|U_{e3}|$ will be observed to be 
close to its current limit, scenarios in which a CKM-like 
$U_\nu$ corrects $U_\ell$ corresponding to $|U_{e3}|=0$ and 
$\theta_{23}= \pi/4$ will be ruled out.\\ 

%

The rephasing invariants 
associated with the Majorana CP violation are given by 
\bea
S_1 \simeq -\cos \theta_{12}^\ell \, 
\left( 
\cos \theta_{12}^\ell \, \sin (\omega + \xi) \, \lambda_{13} + 
\sin \theta_{12}^\ell \, \sin (\omega + \sigma) \, \lambda_{23}\right)
~,\\[0.2cm] 
S_2 \simeq -\sin \theta_{12}^\ell \, 
\left( 
\cos \theta_{12}^\ell \, \sin (\omega - \phi + \xi - \sigma) 
\, \lambda_{13} + 
\sin \theta_{12}^\ell \, \sin (\omega - \phi) \, \lambda_{23}
\right)~.\\[0.2cm] 
\eea
%
In the case of $\lambda_{23}$-dominance 
($\lambda_{13}$-dominance) 
we find that $\beta$ is associated with $\omega + \sigma$ 
$(\omega + \xi)$. In both cases $\alpha$ is 
associated with $\phi + \sigma$.\\

Finally, we give the formulas for the case of a 
CKM-like $U_\nu$, i.e., 
$\lambda_{23} = A \, \lambda_{12}^2$ and 
$\lambda_{13} = B \, \lambda_{12}^3$ with $A$ and $B$ of order one:
\bea
\sin^2 \theta_{12} \simeq \sin^2 \theta_{12}^\ell - \cos \sigma \, 
\sin 2 \theta_{12}^\ell \, \lambda_{12} + \cos 2 \theta_{12}^\ell 
\, \lambda_{12}^2~,\\[0.2cm]
|U_{e3}| \simeq B \, \sin \theta_{12}^\ell \, \lambda_{12}^2~,\\[0.2cm]
\sin^2 \theta_{23} \simeq \frac 12 + B \, \cos \theta_{12}^\ell 
\, \cos(\xi - \sigma + \tau) \, \lambda_{12}^2~,\\[0.2cm]
J_{\rm CP} \simeq -\frac 14 \, B \, \sin 2 \theta_{12}^\ell 
\, \sin \theta_{12}^\ell \, \sin (\xi - \sigma + \tau) 
\, \lambda_{12}^2~.
\eea
%
We note that for an identical in magnitude correction,  
$|U_{e3}|$ is smaller by 
one order in $\lambda_{12}$, 
i.e., $|U_{e3}|\propto \lambda^2_{12}$ 
if the correction comes from $U_\nu$ in contrast to 
$|U_{e3}|\propto \lambda_{12}$ if the correction comes 
from $U_\ell$. 

Consider next the case of $U_\nu$ (and not $U_\nu^\dagger$ as 
before) given by Eq.~(\ref{eq:Unu}). For the neutrino oscillation 
observables we obtain 
\bea
\label{eq:osc2b}
\sin^2 \theta_{12} \simeq \sin^2 \theta_{12}^\ell +  
\lambda_{12} \, \sin 2 \theta_{12}^\ell \, \cos \phi   + 
\frac 14 \, \left( \lambda_{13}^2 - \lambda_{23}^2 \right) \, 
\sin^2 2 \theta_{12}^\ell 
 + \lambda_{12}^2 \, \cos 2 \theta_{12}^\ell  
~,\\[0.2cm]
|U_{e3}| \simeq \left|\lambda_{23} \, \sin \theta_{12}^\ell  
 + \lambda_{13} \, \cos \theta_{12}^\ell \,  
\, e^{i(\phi + \xi)} \right|~,\\[0.2cm]
\sin^2 \theta_{23} \simeq 
\frac 12 \, - \lambda_{23} \, \cos \theta_{12}^\ell 
\, \cos(\omega - \phi) 
 + \lambda_{13} \, 
\sin \theta_{12}^\ell \, \cos (\omega + \xi)
~,\\[0.2cm]
J_{\rm CP} \simeq 
-\frac 14 \, \sin 2 \theta_{12}^\ell \, \left(\lambda_{23} \,  
\sin \theta_{12}^\ell \, \sin(\omega - \phi)  
+ \lambda_{13} \, \cos \theta_{12}^\ell \, \sin (\omega + \xi)  
\right)
~.
\eea 
%
The resulting formulas are very similar to those derived earlier: 
they can be obtained formally from Eq.~(\ref{eq:osc2a})
by simple changes of phases. Since in addition  
$\lambda_{13}$ and $\lambda_{23}$ 
in the expressions for $\sin^2 \theta_{23}$ and $J_{\rm CP}$ 
in Eq.~(\ref{eq:osc2b}) are multiplied by the sines or 
cosines of the same phases, both the sum-rule 
corresponding to $\lambda_{23}$-dominance, Eq.~(\ref{eq:sum2}), 
and the sum-rule associated with $\lambda_{13}$-dominance, 
Eq.~(\ref{eq:sum3}), are valid in this case as well. 


\section{\label{sec:concl}Summary}


The results from various neutrino oscillation experiments indicate 
that $\theta_{23}$ is very close to $\pi/4$ and $\theta_{13}$ is very 
close to zero. It is natural to assume that at 
leading order these mixing 
angles take the quoted extreme values 
and some form of perturbation leads to 
non-zero $\theta_{13}$ and non-maximal $\theta_{23}$. It is hoped 
that this perturbation is imprinted in 
correlations between various observables. 
Future precision experiments can tell us whether there 
are such correlations, which can then be used to 
identify the perturbation and to obtain thereby valuable hints on 
the flavor structure of the underlying theory. In this paper we 
have studied one interesting class of perturbations: because 
the observable lepton mixing matrix is a product of the diagonalization 
matrices of the charged lepton and neutrino mass matrices, 
$U = U_\ell^\dagger \, U_\nu$, we assumed that in the limit of 
one of these matrices being the unit matrix, maximal $\theta_{23}$ and 
zero $\theta_{13}$ would result. When the second matrix deviates 
from being the unit matrix, i.e., has a CKM-like form, we investigated 
the effects on the CP conserving and CP violating observables. 
Free parameters are the small angles of 
the ``correction matrix'', the 12-mixing 
angle of the leading matrix, and various phases. Scenarios like bimaximal 
mixing, tri-bimaximal mixing or Quark-Lepton Complementarity are special 
cases of our analysis. We consistently worked only with rephasing 
invariants in order to avoid the subtleties of identifying CP phases 
within different parameterizations.  
We should stress here also that our analysis is independent 
of the neutrino mass ordering and hierarchy. 

In the first 
scenario we have considered, the neutrino sector alone 
is responsible for zero 
$\theta_{13}$ and maximal $\theta_{23}$. 
Requiring the neutrino mass matrix to obey a 
$\mu$--$\tau$ symmetry can generate such a mixing pattern. 
Figures~\ref{fig:obs1} and \ref{fig:obs1a} 
illustrate the results. We find that $|U_{e3}|$ will typically 
be non-zero, proportional to the sine of the largest angle 
in $U_\ell$, and in most of the cases will be 
well within reach of up-coming experiments. 
If $U_\nu$ is bimaximal,
$|U_{e3}|$ should satisfy $|U_{e3}| \gs 0.1$ in order
for $\sin^2\theta_{12}$ to be within the
$3 \, \sigma$ interval allowed by the current data. 
There is no similar constraint on $|U_{e3}|$
in the case of tri-bimaximal $U_\nu$: even
a vanishing value of $|U_{e3}|$ is allowed.
Atmospheric neutrino mixing can be maximal. There is 
a correlation between the solar neutrino 
mixing, the magnitude of $|U_{e3}|$ and CP violation in 
neutrino oscillations, given by 
$\sin^2 \theta_{12} = \sin^2 \theta_{12}^\nu \pm 
\sqrt{|U_{e3}|^2 \, \sin^2 2 \theta_{12}^\nu - 16 \, J_{\rm CP}^2}$, 
where $\theta_{12}^\nu$ is the 12-rotation angle in $U_\nu$. 
The magnitude of leptonic CP violation is rather sensitive 
to $\theta_{12}^\nu$. 
We have shown as well that
in the approach we are following the 
resulting Dirac CP violating phase, which 
is the source of CP violation 
in neutrino oscillations, cannot 
always be directly identified 
with the Dirac CP violating phase
of the standard PDG parametrization 
of the PMNS matrix. The identification 
of the Majorana CP violating 
phases is typically rather straightforward.

The alternative possibility corresponds to 
the charged lepton sector 
alone being responsible for zero $\theta_{13}$ and 
maximal $\theta_{23}$. We have identified the required 
texture of the charged lepton mass matrix in Eq.~(\ref{eq:mlep}) and plot 
the observables in Fig.~\ref{fig:obs2}. Typically, $|U_{e3}|$ is 
smaller than in the first scenario, being proportional to the sine of 
the second largest angle in $U_\nu$. Another important 
difference with the first case 
is that now there exists a correlation between 
atmospheric neutrino mixing, the magnitude of $|U_{e3}|$ 
and CP violation in neutrino oscillations: 
with $\theta_{12}^\ell$ being the 12-rotation angle in $U_\ell$ 
we find that 
$\sin^2 \theta_{23} \simeq \frac 12 \pm 
\frac{1}{\sin^2 \theta_{12}^\ell} \, 
\sqrt{|U_{e3}|^2 \, \sin^2 \theta_{12}^\ell 
\cos^2 \theta_{12}^\ell - 4 \, J_{\rm CP}^2}$, or 
$\sin^2 \theta_{23} \simeq \frac 12 \mp 
\frac{1}{\cos^2 \theta_{12}^\ell} \, \sqrt{|U_{e3}|^2 \, 
\cos^2 \theta_{12}^\ell \, \sin^2 \theta_{12}^\ell - 4 \, J_{\rm CP}^2}$, 
depending on whether the 23- or 13-rotation angle 
in $U_\ell$ dominates.\\

  We find that both scenarios are in agreement with 
the existing neutrino oscillation data, have interesting 
phenomenology and testable differences.
Future higher precision
determinations of $\sin^2\theta_{12}$
and $\sin^2\theta_{23}$, and 
more stringent constraints on, 
or a measurement of, $|U_{e3}|$ 
can provide crucial tests of these simplest scenarios, 
shedding more light on whether 
any of the two scenarios is 
realized in Nature.

\vspace{0.5cm}
\begin{center}
{\bf Acknowledgments}
\end{center}
This work was supported in part by the Italian MIUR and INFN
programs ``Fundamental Constituents of the Universe'' and 
``Astroparticle Physics'' as well as by the 
Italian MIUR (Internazionalizzazione Program)
and the Yukawa Institute of Theoretical Physics 
(YITP), Kyoto, Japan, within the joint SISSA--YITP 
research project on ``Fundamental Interactions 
and the Early Universe'' (S.T.P.). 
Support by the European Union under 
the ILIAS project, contract No.~RII3--CT--2004--506222 (K.A.H.), 
by the ``Deutsche Forschungsgemeinschaft'' 
in the ``Transregio Sonderforschungsbereich TR 27: 
Neutrinos and Beyond'' (K.A.H.~and W.R.) and under project 
number RO--2516/3--2 (W.R.) is gratefully acknowledged. 
K.A.H.~wishes to thank the 
Max--Planck--Institut f\"ur Kernphysik for its hospitality.


\begin{thebibliography}{99} 

\bibitem{reviews} R.~N.~Mohapatra {\it et al.},
hep-ph/0510213; 
 S.~T.~Petcov,
  Nucl.\ Phys.\ Proc.\ Suppl.\  {\bf 143}, 159 (2005)
  [hep-ph/0412410]; 
R.~N.~Mohapatra and A.~Y.~Smirnov,
hep-ph/0603118; 
A.~Strumia and F.~Vissani,
  hep-ph/0606054.

\bibitem{thomas}T.~Schwetz, 
  hep-ph/0606060; 
  M.~Maltoni {\it et al.}, 
  hep-ph/0405172v5. 

\bibitem{BHP80} S.~M.~Bilenky, J.~Hosek and S.~T.~Petcov,
              Phys. Lett. {\bf B94} 495 (1980).


\bibitem{SchValle80D81} 
 J.~Schechter and J.~W.~F.~Valle,
  Phys.\ Rev.\  D {\bf 22}, 2227 (1980); 
M.~Doi {\it et al.}, 
  Phys.\ Lett.\  B {\bf 102}, 323 (1981).



\bibitem{mutau}An incomplete list of references is 
T.~Fukuyama and H.~Nishiura, hep-ph/9702253; 
R.~N.~Mohapatra and S.~Nussinov, 
Phys.\ Rev.\ D {\bf 60}, 013002 (1999); 
E.~Ma and M.~Raidal, Phys. Rev. Lett. {\bf 87}, 011802 (2001); 
C.~S.~Lam, Phys.\ Lett.\ B {\bf 507}, 214 (2001); 
P.F. Harrison and W. G. Scott, 
Phys.\ Lett.\ B {\bf 547}, 219 (2002); 
T.~Kitabayashi and M.~Yasue, 
Phys.\ Rev.\ D {\bf 67}, 015006 (2003); 
W.~Grimus and L.~Lavoura, 
Phys.\ Lett.\ B {\bf 572}, 189 (2003); 
J.\ Phys.\ G {\bf 30}, 73 (2004); 
Y.~Koide, 
Phys.\ Rev.\ D {\bf 69}, 093001 (2004); 
A. Ghosal, hep-ph/0304090; 
W.~Grimus, A.~S.~Joshipura, S.~Kaneko,
L~.Lavoura, H.~Sawanaka, M.~Tanimoto, 
Nucl.\ Phys.\ B {\bf 713}, 151 (2005); 
R.~N.~Mohapatra, 
JHEP {\bf 0410}, 027 (2004); 
A.~de Gouvea, Phys.\ Rev.\ D {\bf 69}, 093007 (2004); 
S.~Choubey and W.~Rodejohann,
  Eur.\ Phys.\ J.\ C {\bf 40}, 259 (2005); 
 R.~N.~Mohapatra and W.~Rodejohann,
  Phys.\ Rev.\ D {\bf 72}, 053001 (2005); 
 R.~N.~Mohapatra and S.~Nasri, 
  Phys.\ Rev.\ D {\bf 71}, 033001 (2005); 
R.~N.~Mohapatra, S.~Nasri and H.~B.~Yu, 
Phys.\ Lett.\ B {\bf 615}, 231 (2005); 
  Phys.\ Rev.\ D {\bf 72}, 033007 (2005); 
Y.~H.~Ahn {\it et al.}, 
  Phys.\ Rev.\ D {\bf 73}, 093005 (2006); 
hep-ph/0610007; 
T.~Ota and W.~Rodejohann,
  Phys.\ Lett.\ B {\bf 639}, 322 (2006); 
K.~Fuki, M.~Yasue, 
hep-ph/0608042; 
B.~Brahmachari and S.~Choubey,
  Phys.\ Lett.\ B {\bf 642}, 495 (2006); 
 W.~Grimus and L.~Lavoura,
  hep-ph/0611149.




\bibitem{bima}
F.~Vissani, hep-ph/9708483; 
V.~D.~Barger, S.~Pakvasa, T.~J.~Weiler and K.~Whisnant, 
Phys.\ Lett.\ B {\bf 437}, 107 (1998); 
A.~J.~Baltz, A.~S.~Goldhaber and M.~Goldhaber, 
Phys.\ Rev.\ Lett.\  {\bf 81}, 5730 (1998); 
H.~Georgi and S.~L.~Glashow,
Phys.\ Rev.\ D {\bf 61}, 097301 (2000); 
I.~Stancu and D.~V.~Ahluwalia,
Phys.\ Lett.\ B {\bf 460}, 431 (1999). 

\bibitem{tri}P.~F.~Harrison, D.~H.~Perkins and W.~G.~Scott,
  Phys.\ Lett.\ B {\bf 530}, 167 (2002); 
  Phys.\ Lett.\ B {\bf 535}, 163 (2002); 
Z.~Z.~Xing,
  Phys.\ Lett.\ B {\bf 533}, 85 (2002); 
  X.~G.~He and A.~Zee,
  Phys.\ Lett.\ B {\bf 560}, 87 (2003); 
see also 
L.~Wolfenstein,
  Phys.\ Rev.\ D {\bf 18}, 958 (1978).

\bibitem{BCGPRKL2} A.~Bandyopadhyay {\it et al.}, 
{ Phys.\ Lett.} B {\bf 608} 115 (2005), and 2005 (unpublished).


\bibitem{bima_cor}M.~Jezabek and Y.~Sumino,
  Phys.\ Lett.\ B {\bf 457}, 139 (1999); 
Z.~Z.~Xing,
  Phys.\ Rev.\ D {\bf 64}, 093013 (2001); 
  C.~Giunti and M.~Tanimoto,
  Phys.\ Rev.\ D {\bf 66}, 053013 (2002); 
  Phys.\ Rev.\ D {\bf 66}, 113006 (2002); 
W.~Rodejohann,
  Phys.\ Rev.\ D {\bf 69}, 033005 (2004); 
N.~Li and B.~Q.~Ma,
  Phys.\ Lett.\ B {\bf 600}, 248 (2004); 
T.~Ohlsson,
  Phys.\ Lett.\ B {\bf 622}, 159 (2005). 


\bibitem{FPR}P.~H.~Frampton, S.~T.~Petcov and W.~Rodejohann,
  Nucl.\ Phys.\ B {\bf 687}, 31 (2004). 


\bibitem{alta}
G.~Altarelli, F.~Feruglio and I.~Masina,
  Nucl.\ Phys.\ B {\bf 689}, 157 (2004). 

\bibitem{andrea}A.~Romanino,
  Phys.\ Rev.\ D {\bf 70}, 013003 (2004).

\bibitem{AK}
 S.~F.~King, 
  JHEP {\bf 0508}, 105 (2005); 
I.~Masina,
  Phys.\ Lett.\  B {\bf 633}, 134 (2006); 
S.~Antusch and S.~F.~King,
  Phys.\ Lett.\ B {\bf 631}, 42 (2005);  
S.~Antusch, P.~Huber, S.~F.~King and T.~Schwetz,
  JHEP {\bf 0704}, 060 (2007). 

\bibitem{tri_cor}A.~Zee,
  Phys.\ Rev.\ D {\bf 68}, 093002 (2003); 
N.~Li and B.~Q.~Ma,
  Phys.\ Rev.\ D {\bf 71}, 017302 (2005). 
Z.~Z.~Xing,
  Phys.\ Lett.\ B {\bf 533}, 85 (2002).


\bibitem{PlR}
F.~Plentinger and W.~Rodejohann,
  Phys.\ Lett.\ B {\bf 625}, 264 (2005). 




\bibitem{QLC0}M.~Raidal,
  Phys.\ Rev.\ Lett.\  {\bf 93}, 161801 (2004); 
H.~Minakata and A.~Y.~Smirnov,
  Phys.\ Rev.\ D {\bf 70}, 073009 (2004).

\bibitem{QLC1}P.~H.~Frampton and R.~N.~Mohapatra,
  JHEP {\bf 0501}, 025 (2005); 
J.~Ferrandis and S.~Pakvasa,
  Phys.\ Lett.\ B {\bf 603}, 184 (2004); 
S.~K.~Kang, C.~S.~Kim and J.~Lee,
  Phys.\ Lett.\ B {\bf 619}, 129 (2005); 
K.~Cheung {\it et al.}, 
  Phys.\ Rev.\ D {\bf 72}, 036003 (2005); 
A.~Datta, L.~Everett and P.~Ramond,
  Phys.\ Lett.\ B {\bf 620}, 42 (2005); 
S.~Antusch, S.~F.~King and R.~N.~Mohapatra,
  Phys.\ Lett.\ B {\bf 618}, 150 (2005); 
L.~L.~Everett,
  Phys.\ Rev.\ D {\bf 73}, 013011 (2006);  
A.~Dighe, S.~Goswami and P.~Roy,
  Phys.\ Rev.\ D {\bf 73}, 071301 (2006); 
arXiv:0704.3735 [hep-ph]; 
B.~C.~Chauhan {\it et al.}, 
  Eur.\ Phys.\ J.\  C {\bf 50}, 573 (2007); 
F.~Gonzalez-Canales and A.~Mondragon,
  hep-ph/0606175; 
M.~A.~Schmidt and A.~Y.~Smirnov,
  Phys.\ Rev.\  D {\bf 74}, 113003 (2006); 
M.~Picariello,
  hep-ph/0703301.


\bibitem{HR}K.~A.~Hochmuth and W.~Rodejohann,
  Phys.\ Rev.\  D {\bf 75}, 073001 (2007). 

\bibitem{SPAS94} S.~T.~Petcov and A.~Yu.~Smirnov, 
Phys.\ Lett.\ B {\bf 322}, 109 (1994).


\bibitem{PR}S.~T.~Petcov and W.~Rodejohann,
Phys.\ Rev.\ D {\bf 71}, 073002 (2005).

\bibitem{lelmlt}S.~T.~Petcov,
  Phys.\ Lett.\ B {\bf 110} 245 (1982). 

\bibitem{mutaulep0}Z.~Z.~Xing,
  Phys.\ Lett.\ B {\bf 618}, 141 (2005). 

\bibitem{mutaulep1}R.~N.~Mohapatra and W.~Rodejohann,
  Phys.\ Rev.\ D {\bf 72}, 053001 (2005). 


\bibitem{Pi}M.~Picariello,
  hep-ph/0611189.

\bibitem{anki0}S.~Antusch and S.~F.~King,
  Phys.\ Lett.\ B {\bf 591}, 104 (2004).


\bibitem{others}C.~A.~de S. Pires,
  J.\ Phys.\ G {\bf 30}, B29 (2004); 
N.~Li and B.~Q.~Ma,
  Eur.\ Phys.\ J.\ C {\bf 42}, 17 (2005); 
T.~Ohlsson,
  Phys.\ Lett.\ B {\bf 622}, 159 (2005), see also 
 F.~Plentinger, G.~Seidl and W.~Winter,
  hep-ph/0612169.


\bibitem{PPR03} S.~Pascoli, S.~T.~Petcov and W.~Rodejohann, 
Phys.\ Rev.\ D {\bf 68}, 093007 (2003).

\bibitem{STPNob04} 
S.~T.~Petcov,
  Phys.\ Scripta {\bf T121}, 94 (2005)
  [hep-ph/0504166].


\bibitem{BPP1}
S.~M.~Bilenky, S.~Pascoli and S.~T.~Petcov,
  Phys.\ Rev.\  D {\bf 64}, 053010 (2001). 


\bibitem{PKSP3nu88}
P.~I.~Krastev and S.~T.~Petcov,
  Phys.\ Lett.\  B {\bf 205}, 84 (1988).


\bibitem{CJ85} C.~Jarlskog, 
{Z.\ Phys.} C {\bf 29} 491 (1985); 
Phys.\ Rev.\ Lett.\  {\bf 55}, 1039 (1985).







\bibitem{JMaj87} 
J.~F.~Nieves and P.~B.~Pal, 
Phys.\ Rev.\ D {\bf 36}, 315 (1987); 
Phys.\ Rev.\ D {\bf 64}, 076005 (2001).

\bibitem{ASBranco00} J.~A.~Aguilar-Saavedra and G.~C.~Branco, 
Phys.\ Rev.\ D {\bf 62}, 096009 (2000).  


\end{thebibliography}
\end{document}